\documentclass[pra,aps,superscriptaddress,superbib,citeautoscript,reprint]{revtex4-1}

\usepackage[english]{babel}
\usepackage[utf8]{inputenc}
\usepackage[T1]{fontenc}

\usepackage[colorlinks=true,urlcolor=blue,citecolor=blue,linkcolor=blue]{hyperref}

\usepackage{amsmath,bm}
\usepackage{amstext}
\usepackage{epsfig}
\usepackage{xcolor}
\usepackage{graphicx,amsmath,amssymb,tabularx}
\usepackage{multirow}
\usepackage{array}
\usepackage{dsfont}
\usepackage{cleveref}
\usepackage[toc]{appendix}
\usepackage{physics}
\usepackage{mathrsfs} 

\definecolor{dgreen}{rgb}{0,.5,0}
\definecolor{dred}{rgb}{.7,.0,.0}

\newcommand{\etal}{{\it et al.}}

\DeclareMathAlphabet\mathbfcal{OMS}{cmsy}{b}{n}

\newcommand{\bxi}{\bm{\xi}}

\newcommand{\br}{\mathbf{r}}

\newcommand{\tw}{{\tt w}}
\newcommand{\ie}{{\it i.e.}}
\newcommand{\myket}[1]{\left\vert #1\right\rangle}
\newcommand{\mybra}[1]{\left\langle #1\right\vert}
\newcommand{\bmeta}{{\bm \eta}}

\newcommand{\dxi}[1]{\dfrac{\partial #1}{\partial \xi}}
\newcommand{\dxim}[1]{\dfrac{\partial #1}{\partial \xi_-}}
\newcommand{\dxip}[1]{\dfrac{\partial #1}{\partial \xi_+}}
\newcommand{\dw}[1]{\dfrac{\partial #1}{\partial \tw}}
\newcommand{\ximu}{\xi_{\mu}}
\newcommand{\xinu}{\xi_{\nu}}
\newcommand{\xilambda}{\xi_{\lambda}}

\newcommand{\dxilambda}[1]{\dfrac{\partial #1}{\partial \xi_\lambda}}

\newcommand{\bruno}[1]{{\textcolor{blue}{ Bruno: #1 }} }

\newcommand{\be}{\begin{equation}}
\newcommand{\ee}{\end{equation}}
\DeclareMathAlphabet\mathbfcal{OMS}{cmsy}{b}{n}

\newcommand{\rev}[1]{{{#1}}}

\newcommand{\titou}[1]{\textcolor{red}{#1}}

\newcommand{\PFL}[1]{\titou{(\underline{\bf PFL}: #1)}}

\newcommand{\LCPQ}{Laboratoire de Chimie et Physique Quantiques (UMR 5626), Universit\'e de Toulouse, CNRS, UPS, France}
\newcommand{\LCQ}{Laboratoire de Chimie Quantique, Institut de Chimie, CNRS/Universit\'{e} de Strasbourg, 4 rue Blaise Pascal, 67000 Strasbourg, France}
\newcommand{\ICGM}{ICGM, Université de Montpellier, CNRS, ENSCM, Montpellier, France}

\begin{document}

\title{
Neutral electronic excitations and derivative discontinuities: An
extended $N$-centered ensemble density functional theory perspective 
}

\author{Filip Cernatic}
\affiliation{\LCQ}
\author{Pierre-Fran\c{c}ois Loos}
\affiliation{\LCPQ}
\author{Bruno Senjean}
\affiliation{\ICGM}
\author{Emmanuel Fromager}
\affiliation{\LCQ}


\begin{abstract}
This work merges two different types of many-electron ensembles,
namely the Theophilou--Gross--Oliveira--Kohn ensembles of ground and
neutrally-excited states, and the more recent $N$-centered ensembles of
neutral and charged ground states. On that basis, an in-principle exact
and general, so-called extended $N$-centered, ensemble density-functional 
theory of charged and neutral electronic excitations is derived. 
We revisit in this context the concept of density-functional
derivative discontinuity for neutral excitations, without ever invoking
nor using the asymptotic behavior of the ensemble electronic density.
The present mathematical construction fully relies on the 
weight dependence of the ensemble Hartree-exchange-correlation density-functional
energy, which makes the theory applicable to lattice models and
opens new perspectives for the description of gaps in mesoscopic systems.  
\end{abstract}

\maketitle



\section{Introduction}

Kohn--Sham density-functional theory (KS-DFT)~\cite{KS} is prominently used nowadays to obtain approximate but reliable ground-state electronic energies around the equilibrium geometries of large systems with up to thousands of electrons.
To go beyond ground-state properties, time-dependent
extensions have been proposed, such as
linear response time-dependent DFT (TD-DFT),
where the first-order KS response function and the Hartree-exchange-correlation (Hxc) 
density-functional kernel allow for the extraction of excited-state energies~\cite{Casida_tddft_review_2012},
or real-time (or propagation) TD-DFT that relies on the numerical propagation of the electronic equations-of-motion and is not limited in principle to perturbations falling into the linear regime~\cite{runge1984density}.
Alternatively, and popularized in condensed matter physics, the more involved many-body perturbation theory \cite{onida2002electronic,MartinBook} within $GW$ \cite{hedin1965new,golze2019GW,marie2023GW} together with the Bethe--Salpeter equation, which relies on the two-particle (frequency-dependent) Green’s function, can be used~\cite{strinati1988application,sottile2007efficient,blase2020bethe}.
All the aforementioned methods are computationally more demanding than KS-DFT.
Even more problematic, linear response TD-DFT does not generally give an
accurate description of charge transfer
excitations~\cite{fuks2014challenging}, and multiple excitations are
even absent from the spectra~\cite{maitra2004double}, when the popular
semi-local adiabatic approximation
is employed~\cite{Casida_tddft_review_2012,Lacombe2023_Non-adiabatic}.

During the last decade, increasing attention has been paid to
time-independent ensemble extensions of DFT (ensemble
DFT)~\cite{PRA13_Pernal_srEDFT,pastorczak2014ensemble,franck2014generalised,yang2014exact,pribramjones2014excitations,filatov2015spin,senjean2015linear,deur2017exact,yang2017direct,gould2017hartree,gould2018charge,deur2018exploring,PRL19_Gould_DD_correlation,Fromager_2020,PRL20_Gould_Hartree_def_from_ACDF_th,Gould2020_Approximately,loos2020weightdependent,Gould2021_Ensemble_ugly,gould2023local,Gould2023_Electronic,gould2022single,Gould2021_Double,Cernatic2022}, from which in-principle exact excitation energies can be inferred with the same computational cost as regular (ground-state) KS-DFT.
Such an extension was originally proposed
by Theophilou~\cite{JPC79_Theophilou_equi-ensembles,theophilou_book},
Gross, Oliveira, and Kohn~\cite{gross1988density}
for {\it neutral} excitation energies (\ie, difference of energy between states of a system with $N$ electrons)
and is referred to as TGOK-DFT, following the name of the authors. The
use of TGOK ensembles in other settings than DFT,
both for the description of electronic and
bosonic low-lying excited states, has also become increasingly appealing~\cite{Schilling2021_Ensemble,Liebert2022_Foundation,Benavides-Riveros2022_Excitations,Liebert_2023_An_exact_bosons,Liebert2023_Deriving}. 

Quite recently, the concept of $N$-centered ensemble~\cite{senjean2018unified,Senjean_2020} has been introduced
by analogy with TGOK-DFT to evaluate, in principle exactly,
{\it charged} excitation energies
(\ie, excitation energies between ground states corresponding to different electron numbers). 
Importantly, $N$-centered ensemble DFT
shed a new light on the concept of density-functional derivative
discontinuity for charged excitations,
which emerged from the seminal work of Perdew, Parr, Levy, and Balduz (PPLB) in the context of DFT for fractional particle number~\cite{perdew1982density,perdew1983physical}.
Within the formulation of PPLB, it is in principle sufficient to extend the
domain of definition of the Hxc density functional to fractional electron numbers in order to account
for derivative discontinuities.
However, it has been argued that invoking a fractional number
of electrons is maybe not the correct route 
to pursue~\cite{baerends2013kohn,
baerends2017kohn,baerends2018density,Baerends2022_Chemical,Baerends2020_On_derivatives}.
This debate aligns with the approach taken in $N$-centered ensemble
DFT, where the derivative discontinuity is alternatively described
through weight derivatives of the ensemble density-functional Hxc energy at {\it fixed} ensemble density.
Such derivatives are well defined in the $N$-centered formalism since, by
construction, the ensemble density always integrates to the fixed and central (integer)
number of electrons $N$, hence the name of the theory.
Extensive discussions about the equivalence between weight derivatives
and derivative discontinuities for ground states
can be found in Ref.~\cite{Cernatic2022}.

While a regular (so-called \textit{left} in Ref.~\onlinecite{Senjean_2020}) $N$-centered	
ensemble consists of $(N-1)$- and $N$-electron ground states,
we propose in the present work to incorporate 
neutrally-excited (\ie, $N$-electron excited) states into the ensemble, thus
allowing, for example, to decompose neutral excitation processes into
separate ionization 
ones, as originally suggested by Levy~\cite{levy1995excitation}. 
\rev{On that basis, exactifying KS orbital energies in the description of
neutral excitations consists in exactifying both ground- and
excited-state KS ionization potentials, a procedure from which derivative
discontinuities emerge~\cite{levy1995excitation}. As mentioned
previously, in $N$-centered ensemble DFT, the description of 
derivative discontinuities (that connect to the piecewise-linear
behavior of the ground-state
energy~\cite{perdew1982density,perdew1983physical,kraisler2013piecewise})
is recast into the modeling of weight dependencies in the ensemble Hxc
density functional~\cite{PRA21_Hodgson_exact_Nc-eDFT_1D,Cernatic2022}.
Even though its practical advantage still needs to be assessed, in
particular because the incorporation of weight dependencies 
into regular density-functional approximations is not straightforward
when the ensemble energy is evaluated
variationally~\cite{Gould2021_Ensemble_ugly,Cernatic2022}, the $N$-centered strategy offers  
a mathematically elegant and simplified alternative in describing how KS
orbital energies are affected by charged excitation processes. Our
motivation is to extend this strategy to neutral
excitations.}
Even though we do not exploit in this work the following
possibility, the formalism is flexible enough to include anionic states. The
resulting \rev{combined} TGOK/$N$-centered ensemble formalism, that we refer to as {\it extended}
$N$-centered, lays the foundations of a unified and general
ensemble density functional theory of charged and neutral electronic
excitations. Most importantly, as shown in the following, this formalism
sheds a different light on the concept of
derivative discontinuity for {\it neutral}
excitations~\cite{levy1995excitation,yang2014exact,Kraisler21_From,gould2022single},
which is much less discussed
in the literature than for charged excitations.

The paper is organized as follows.
After motivating in Sec.~\ref{sec:problematic} the decomposition of a
neutral excitation process into
two ionization processes, and reviewing briefly TGOK-DFT in Sec.~\ref{sec:review}, we introduce the extended $N$-centered ensemble
density-functional formalism in Sec.~\ref{sec:eNc_ensemble_DFT}. On that
basis, the concept of derivative discontinuity in the context of neutral
excitation processes is revisited in Sec.~\ref{sec:DDs_revisited}. 
The theory is then applied to the two-electron Hubbard dimer, as a
proof of concept. The exact derivations and numerical tests are presented in Sec.~\ref{sec:Application_Hdim}.
Conclusions and outlook are finally given in Sec.~\ref{sec:conclu}.

\section{Problematic: On the exactification of Kohn--Sham orbital energies}\label{sec:problematic}

The occupied and virtual orbital energies
$\left\{\varepsilon_i\right\}_{i=1,2,\ldots}$ generated from a regular
$N$-electron ground-state KS-DFT calculation can be used to compute
total ground- and excited-state $N$-electron KS energies, 
\be
\mathcal{E}^N_{\nu}=\sum_i n_\nu^i\varepsilon_i,
\ee
where $n_\nu^i\in\left\{0,1,2\right\}$ denotes the integer occupation of
the $i$th KS orbital in the $\nu$th KS state ($\nu\geq 0$) and $\sum_i
n^i_\nu=N$. It is well-known that when it comes to describing neutral
excitation processes, the bare KS excitation energies 
and the true
interacting ones $\left\{\omega_\nu\right\}_{\nu>0}$ do not match:  
\be
\omega_\nu\equiv
E^N_\nu-E^N_0 \neq \mathcal{E}^N_{\nu}-\mathcal{E}^N_{0}.
\ee

In the context of linear response TD-DFT, these two quantities are
connected through the Hxc kernel, {\it i.e.}, the
density-functional derivative of the (time-dependent) Hxc potential. We
focus in the following on in-principle-exact time-independent density-functional approaches to neutral
electronic excitations and, more specifically, to ensemble ones. At this
point we should stress that, unlike in charged processes, any constant
shift in the Hxc potential and therefore in the orbital energies,    
\be
\varepsilon_i\rightarrow \varepsilon_i+c,
\ee
leaves neutral KS excitation energies unchanged: 
\be\label{eq:shift_pot_no_change_KS_XE}
\mathcal{E}_{\nu}^N-\mathcal{E}^N_{0}\rightarrow
\left(\mathcal{E}^N_{\nu}+Nc\right)-\left(\mathcal{E}^N_{0}+Nc\right)
=\mathcal{E}^N_{\nu}-\mathcal{E}^N_{0}.\ee
From this standpoint, it seems impossible to exactify the KS orbital
energies in the description of neutrally excited states. Nevertheless,
as recalled in Sec.~\ref{sec:review}, it is possible to describe exactly the
deviation of the physical excitation energy from the KS one by means of
an ensemble-weight-dependent Hxc density functional.
\rev{A subtle but crucial point, which has not been taken into account in
Eq.~(\ref{eq:shift_pot_no_change_KS_XE}), is the fact that the KS
orbital energies of the reference $N$-electron ground-state KS-DFT
calculation are not
those that describe the (even infinitesimally) excited system~\cite{Kraisler21_From}. Interestingly, a similar issue has already been discussed for the
fundamental gap problem in the context of $N$-centered ensemble
DFT~\cite{PRA21_Hodgson_exact_Nc-eDFT_1D}
where, by construction, the ensemble Hxc potential is also defined up to a
constant, unlike in the more conventional DFT for fractional electron
numbers~\cite{Cernatic2022}. In fact, within $N$-centered ensemble
DFT, the fundamental and optical gap problems have the exact same
mathematical formulation~\cite{senjean2018unified}. What
Ref.~\citenum{PRA21_Hodgson_exact_Nc-eDFT_1D} revealed (in 1D) is that, if the
$N$-centered ensemble Hxc potential is (arbitrarily) adjusted such that
it asymptotically approaches zero infinitely far from the center of the
system under study, by plotting the difference in Hxc potentials between, on the one
hand, the ensemble
that contains both ground $N$-electron and
$(N+1)$-electron (with weight $\xi_+$) states, and, on the other hand, the regular pure
$N$-electron ground state ($\xi_+=0$ in this case), one can see a plateau in
the central region of the system that grows in length as $\xi_+\rightarrow
0^+$. Ultimately, it looks as if the regular Hxc potential for $N$
electrons in the ground state had been shifted, thus ensuring, in the
present case, the
exactification of the KS electron affinity when $\xi_+\rightarrow 0^+$
(and of the KS ionization potential when $\xi_+=0$). In summary, the KS orbital energies are
affected by the (even infinitesimal) occupation of higher-energy orbitals and
this is reflected by a shift in the Hxc potential. A similar analysis
has been performed by Kraisler \etal~\cite{Kraisler21_From} for neutral
excitations, which is our main focus.}

In the present work, we follow a connected but
slightly different path. \rev{We propose to achieve
a description of neutral excitations through an exactification of the KS orbital energies that exclusively relies on an appropriate (so-called extended
$N$-centered) ensemble weight-dependent Hxc
density functional, by analogy with the $N$-centered ensemble dealing with the
fundamental gap~\cite{PRA21_Hodgson_exact_Nc-eDFT_1D,Cernatic2022}}. 
Based on the
observation made in
Eq.~(\ref{eq:shift_pot_no_change_KS_XE}) and the seminal work of
Levy~\cite{levy1995excitation}, the key idea consists in evaluating a neutral
excitation energy {\it via} two different {\it charged}
processes, namely the ionization of the ground ($\nu=0$) and the
targeted excited ($\nu>0$)
$N$-electron states, {\ie},               
\be\label{eq:neutral_XE_from_IPs}
E^N_\nu-E^N_0 = I^N_0-I^N_\nu,
\ee
where 
\be\label{eq:IP_N-electron_nu-state}
\left\{I^N_\nu = E^{N-1}_0-E^N_\nu\right\}_{\nu\geq
0}
\ee
will be referred to as ground- and excited-state ionization potentials (IPs). 


Turning to the KS system, for each
ionization process, a specific shift can be applied to the Hxc
potential (which is
unique up to a constant, as long as the number of electrons is fixed to
the integer $N$) in order to {\it enforce} the KS IPs to
match the true interacting ones, \ie, 
\be\label{eq:exact_Izero_from_KS_c1}
\begin{split}
\varepsilon_i&\rightarrow \varepsilon_i+c_1=\overline{\varepsilon}_i,
\\
\mathcal{E}^{N-1}_{0}-\mathcal{E}^N_{0}&\rightarrow
\mathcal{E}^{N-1}_{0}-\mathcal{E}^N_{0}-c_1 \overset{!}{=} I^N_0, 
\end{split}
\ee
and
\be\label{eq:exact_Inu_from_KS_c2}
\begin{split}
\varepsilon_i&\rightarrow \varepsilon_i+c_2=\tilde{\varepsilon}_i,
\\
\mathcal{E}^{N-1}_{0}-\mathcal{E}^N_{\nu}&\rightarrow
\mathcal{E}^{N-1}_{0}-\mathcal{E}^N_{\nu}-c_2 
\overset{!}{=} I^N_\nu. 
\end{split}
\ee
If, for example, the excited state of interest $\nu$ is
described by a single-electron excitation (one hole, one particle) from the highest occupied
molecular orbital (HOMO) $i=N$ to a virtual one \rev{$i(\nu)>N$}, then we
automatically obtain from
Eqs.~(\ref{eq:neutral_XE_from_IPs}), (\ref{eq:exact_Izero_from_KS_c1}),
and (\ref{eq:exact_Inu_from_KS_c2}) what we consider as the exactification of the KS
orbital energies for neutral excitations, {\ie},   
\be\label{eq:separate_shifts_acceptor_and_homo}
E^N_\nu-E^N_0=\rev{\tilde{\varepsilon}_{i(\nu)}}-\overline{\varepsilon}_N.
\ee

The question that is addressed in the rest of this work is how such a
construction can be derived, in principle exactly \rev{(and with no need to invoke the asymptotic behavior of the ensemble density infinitely far from the center of the
system under study, as
will be discussed later)}, from a unified and
general
ensemble density-functional formalism in which both charged and neutral
excitation processes can be described simultaneously. 

\section{Brief review of regular TGOK ensemble DFT}\label{sec:review}

TGOK-DFT is a time-independent ensemble extension of standard ground-state
DFT to neutral excited states where the ground-state energy is replaced
by the so-called ensemble energy, which is a convex combination of ($N$-electron)
ground- and excited-state energies,
\be\label{eq:TGOK_ens_energy}
E^{\bxi}\overset{\rm TGOK}{:=}
\left(1-\sum_{\nu>0}\xi^N_\nu\right)E_0^N+\sum_{\nu>0}\xi^N_\nu
E^N_\nu,
\ee
where 
\be
\bxi
\overset{\rm TGOK}{\equiv}
 \left\{\xi^N_\nu\right\}_{\nu>0}
\ee 
is the collection of
positive and {\it independent} ensemble weight values that are assigned
to the ordered-in-energy 
$N$-electron excited states $\left\{\Psi_\nu^N\right\}_{\nu>0}$, {\ie},
\be\label{eq:TGOK_ordered_energies}
E_\nu^N=\mel{\Psi_\nu^N}{\hat{H}}{\Psi_\nu^N}<E_{\nu+1}^N.
\ee 
The electronic Hamiltonian operator $\hat{H}$ of the system under study reads 
\be
\hat{H}=\hat{T}+\hat{W}_{\rm ee}+\hat{V}_{\rm
ext}
,
\ee
where $\hat{T}$ describes the kinetic
energy, $\hat{W}_{\rm ee}$ is the two-electron
repulsion operator, $\hat{V}_{\rm
ext}=\int d\br\, v_{\rm
ext}(\br)\hat{n}(\br)$ is the external local potential operator, and
$\hat{n}(\br)$ denotes the density operator at position $\br$. Note that
the weight $\xi_0^N$ assigned to the ground-state energy $E_0^N$ in
Eq.~(\ref{eq:TGOK_ens_energy}) is such that the collection of weights
(including the ground-state one) is normalized:
\be
\xi_0^N+\sum_{\nu>0}\xi^N_\nu\overset{\rm{TGOK}}{=}1.
\ee
This constraint, where $\xi_0^N$ is an affine function of the independent
excited-state ensemble weights~\cite{deur2019ground}, ensures that the total number of
electrons is preserved when deviating from the ground-state $\bxi=0$
limit of the theory, {\ie},
\be\label{eq:ens_dens_integrates_to_N}
\int d\br\,n^{\bxi}(\br)=N,\quad \forall \bxi,
\ee   
where
\be
n^{\bxi}(\br)\overset{\rm TGOK}{:=}
\left(1-\sum_{\nu>0}\xi^N_\nu\right)n_{\Psi_0^N}({\br})
+\sum_{\nu>0}\xi^N_\nu
n_{\Psi_\nu^N}({\br})
\ee 
is the ensemble density, $\left\{n_{\Psi_\nu^N}({\br})\right\}_{\nu\geq
0}$ being the individual $N$-electron ground- and excited-state
densities. In the following, we use the more compact notation,
\be\label{eq:ens_dens_from_Gamma_hat}
n^{\bxi}(\br)
=\Tr[\hat{\Gamma}^{\bxi}\hat{n}(\br)],
\ee
where 
\be
\hat{\Gamma}^{\bxi}\overset{\rm TGOK}{=}\qty(1 -
\sum_{\nu>0}\xinu^N)\dyad*{{\Psi}_0^N}
+\sum_{\nu>0}\xinu^N\dyad*{{\Psi}^N_{\nu}}
\ee
is the ensemble density matrix operator and $\Tr$ denotes the trace. Note that we have assumed, for simplicity,
that both ground and excited states are not degenerate but the formalism
can be extended straightforwardly to ensembles of multiplets~\cite{gross1988density}.\\

The TGOK ensemble energy, as defined in Eqs.~(\ref{eq:TGOK_ens_energy})
and (\ref{eq:TGOK_ordered_energies}), can be determined variationally,
{\ie},
\be\label{eq:TGOK_ens_ener_var_principle}
E^{\bxi}=\min_{\hat{\gamma}^{\bxi}}\Tr[\hat{\gamma}^{\bxi}\hat{H}]=\Tr\left[\hat{\Gamma}^{\bxi}\hat{H}\right],
\ee    
where $\hat{\gamma}^{\bxi}$ is a trial ensemble density matrix operator,
provided that the ensemble weights are collected in decreasing
order~\cite{gross1988rayleigh}:
\be\label{TGOK_weights_ordering}
\xi^N_\nu\geq \xi^N_{\nu+1}, \quad \nu\geq 0.
\ee
On that basis, an {ensemble KS-DFT} of neutral excited states can be
derived~\cite{gross1988density}, where the ensemble Hxc density
functional is defined
exactly, as follows
\be\label{eq:ensemble_Hxc_formal_exp}
\begin{split}
E_{\rm Hxc}^{\bxi}[n]
&=\min_{\hat{\gamma}^{\bxi}\rightarrow
n}\Tr[\hat{\gamma}^{\bxi}\left(\hat{T}+\hat{W}_{\rm ee}\right)]
-\min_{\hat{\gamma}^{\bxi}\rightarrow
n}\Tr[\hat{\gamma}^{\bxi}\hat{T}]
\\
&:=F^{\bxi}[n]-T^{\bxi}_{\rm s}[n],
\end{split}
\ee
where $F^{\bxi}[n]$ (and its noninteracting analogue $T^{\bxi}_{\rm
s}[n]$) is simply the extension to ensembles (with given fixed ensemble weight
values $\bxi$) of the Levy--Lieb
functional~\cite{levy1979universal,LFTransform-Lieb}. 
The density constraint used in both minimizations reads
$n_{\hat{\gamma}^{\bxi}}(\br)=\Tr[\hat{\gamma}^{\bxi}\hat{n}(\br)]=n(\br)$.

Note that the ensemble Hxc functional is both a functional of the
density $n$ {\it and} a function of the ensemble
weights $\bxi$. The reason is that, unlike in ground-state DFT of open
electronic systems, where the ensemble weight is deduced from the
fractional number of electrons, a TGOK ensemble cannot be identified solely from its density.
Indeed, a density $n$ that integrates to a given {\it integer} number $N$ of
electrons may be both pure-ground-state and
ensemble $v$-representable at the same time [see Eq.~(37) in
Ref.~\onlinecite{Teale2022_DFT_exchange} and the discussion that
follows]. Without additional information about the
ensemble (namely the ensemble weight values), the functional would not
``know'' if it has to compute the Hxc energy of a pure ground state
($\bxi=0$) or
that of an ensemble ($\bxi>0$). 

Finally, like in ground-state KS-DFT,
the ensemble density-functional Hxc energy can be decomposed into
(weight-dependent) Hx and correlation contributions, which 
read~\cite{gould2017hartree}
\rev{
\begin{equation}
\begin{split}
E_{\rm Hx}^{\bxi}[n] & =\lim_{\alpha\rightarrow
0^+}\dfrac{F^{\bxi,\alpha}[n]-F^{\bxi,\alpha=0}[n]}{\alpha}
\\
&=\left.\dfrac{
\partial F^{\bxi,\alpha}[n]}{\partial \alpha}\right|_{\alpha=0^+},
\end{split}
\end{equation}
and
\be
E_{\rm c}^{\bxi}[n]=E_{\rm Hxc}^{\bxi}[n]-E_{\rm Hx}^{\bxi}[n],
\ee
respectively, where 
\be
F^{\bxi,\alpha}[n]=\min_{\hat{\gamma}^{\bxi}\rightarrow
n}\Tr[\hat{\gamma}^{\bxi}\left(\hat{T}+\alpha\hat{W}_{\rm ee}\right)]
\ee
is the analog for electrons that partially interact (through the
positive scaling $\alpha$ of the electronic
repulsion) of the TGOK Levy--Lieb functional $F^{\bxi}[n]$ introduced in
Eq.~(\ref{eq:ensemble_Hxc_formal_exp}). Note that $F^{\bxi,\alpha=1}[n]=F^{\bxi}[n]$
and $F^{\bxi,\alpha=0}[n]=T^{\bxi}_{\rm s}[n]$.} 
\rev{It is important to stress that, unlike in regular pure ground-state
KS-DFT, neither the definition of the Hx energy nor its separation into Hartree and exchange
contributions are
straightforward in the context of ensemble DFT~\cite{ensemble_ghost_interaction,pastorczak2014ensemble,PRL20_Gould_Hartree_def_from_ACDF_th,Gould2021_Ensemble_ugly,Cernatic2022}}.
The latter point will be discussed a bit
further after Eq.~(\ref{eq:GSH_plus_ens_xc}).\\

In the context of TGOK-DFT, the variational ensemble energy
expression of Eq.~(\ref{eq:TGOK_ens_ener_var_principle}) can be recast into
a minimization over noninteracting ensemble density matrix operators
\be\label{eq:variational_ens_ener_KS_scheme}
E^{\bxi}=\min_{\hat{\gamma}^{\bxi}}\left\{\Tr[\hat{\gamma}^{\bxi}\left(\hat{T}+\hat{V}_{\rm
ext}\right)]+E_{\rm Hxc}^{\bxi}[n_{\hat{\gamma}^{\bxi}}]\right\},
\ee
where the (weight-dependent~\cite{Cernatic2022}) minimizing KS wave
functions $\left\{\Phi^{\bxi}_\nu\right\}_{\nu\geq 0}$ fulfill the
following self-consistent noninteracting Schr\"{o}dinger equation
\be\label{eq:ens_KS_eqs_many-body_KS_states}
\left[\hat{T}+\hat{V}_{\rm
ext}+\int d\br\,v^{\bxi}_{\rm
Hxc}(\br)\hat{n}(\br)\right]\ket{\Phi^{\bxi}_\nu}=\mathcal{E}^{\bxi}_\nu\ket{\Phi^{\bxi}_\nu},
\ee
$v^{\bxi}_{\rm
Hxc}(\br)=\left.\delta E_{\rm Hxc}^{\bxi}[n]/\delta
n(\br)\right|_{n=n^{\bxi}}$ being the weight-dependent Hxc potential.
Equivalently, the orbitals from which the KS ensemble is
constructed fulfill the following ensemble KS equations
\be\label{eq:ens_KS_eqs_orbitals}
\left[-\dfrac{\nabla_{\br}^2}{2}+v_{\rm ext}(\br)+v^{\bxi}_{\rm
Hxc}(\br)\right]\varphi^{\bxi}_i(\br)=\varepsilon^{\bxi}_i\varphi^{\bxi}_i(\br).
\ee
The latter differ from regular (ground-state) KS equations by i) the weight
dependence of the Hxc potential, and ii) the fact that the KS
orbitals, which reproduce the exact ensemble density $n^{\bxi}(\br)$,
are fractionally occupied, \ie,  
\be
\label{eq:ens_dens_from_KS_many_body_states}
n^{\bxi}(\br)=
\sum_{\nu\geq 0}\xi_\nu n_{\Phi^{\bxi}_\nu}(\br)
=\sum_i\left(\sum_{\nu\geq 0}\xi_\nu
n^i_\nu\right)\abs{\varphi^{\bxi}_i(\br)}^2,
\ee
where $n_\nu^i$ is the occupation of the KS orbital $\varphi^{\bxi}_i$
in the KS analog $\Phi^{\bxi}_\nu$ of the pure state $\nu$. Note that
the (weight-dependent) total KS energies simply read 
\be\label{eq:total_KS_ener_from_KS_orb_ener}
\mathcal{E}^{\bxi}_{\nu}=\sum_i n_\nu^i\varepsilon^{\bxi}_i.
\ee  

Turning to the problematic raised in Sec.~\ref{sec:problematic}, we
should first recall that the exact deviation of the true interacting
excitation energies from the KS ones is given by the derivative with
respect to the ensemble weights (at fixed ensemble density) of the ensemble Hxc
density-functional energy~\cite{gross1988density,deur2019ground}:
\be\label{eq:excitation_energy_from_TGOK-DFT}
E^N_\nu-E^N_0-\left(\mathcal{E}^{\bxi}_{\nu}-\mathcal{E}^{\bxi}_{0}\right)=
\left.\dfrac{\partial E_{\rm
Hxc}^{\bxi}[n]}{\partial\xi^N_\nu}\right|_{n=n^{\bxi}}.
\ee
Therefore, within regular TGOK-DFT, the exactification of the ensemble
KS orbital energies occurs only when the above ensemble weight
derivative vanishes, which may happen for very specific ({\it a
priori} unknown) weight values~\cite{yang2014exact,deur2017exact}. In
other words, such an exactification cannot, in general, be achieved, unless we introduce intermediate
ionization processes, as suggested in Sec.~\ref{sec:problematic}. The
main challenge in this case lies in the design of a unified ensemble
density-functional formalism where both neutral and charged excitations
can be described. A solution to this 
problem is proposed in Sec.~\ref{sec:eNc_ensemble_DFT}. 

\rev{As a side comment, which aims at preventing confusions later on, we
want to stress that, in principle, the Hxc ensemble weight derivative
[\ie, the KS excitation energy correction on the right-hand side of
Eq.~(\ref{eq:excitation_energy_from_TGOK-DFT})] and the Hxc
density-functional derivative discontinuity~\cite{yang2014exact,Kraisler21_From,gould2022single} are two distinct concepts. The latter, which affects the energy
of the KS acceptor orbital $i(\nu)$ introduced in
Eq.~(\ref{eq:separate_shifts_acceptor_and_homo}),  
appears in the $\bxi\rightarrow {\bm 0}^+$ limit of the theory, when comparison is made with regular
$N$-electron ground-state KS-DFT (where $\bxi={\bm
0}$) from which the energy of
the HOMO is evaluated. The two concepts become equivalent in this
special case, as it will become clear from the upcoming    
Eqs.~(\ref{eq:eq:Exact_Omega_from_KS_orb_energies_zero_weight_limit})
and (\ref{eq:HxcDD_from_NceDFT_final_result}), once the extended
$N$-centered ensemble DFT formalism has been introduced. This is the
purpose of Sec.~\ref{sec:eNc_ensemble_DFT}.}

\section{Extended $N$-centered ensemble DFT}\label{sec:eNc_ensemble_DFT}

Senjean and Fromager~\cite{senjean2018unified} introduced some years ago the so-called
$N$-centered ensemble DFT formalism where the fundamental gap of
$N$-electron ground states is described with the exact same mathematical
language as in TGOK-DFT. The approach also allows for a separate
description of ionization and affinity
processes~\cite{senjean2018unified,Senjean_2020,PRA21_Hodgson_exact_Nc-eDFT_1D,Cernatic2022}, which is essential in
the present context. The close resemblance of
$N$-centered ensemble DFT with TGOK-DFT is exploited in the following in
order to provide an in-principle exact ensemble density-functional
description of ionized excited states. The resulting ensemble formalism,
where neutral excited states are incorporated into a regular
(ground-state) $N$-centered ensemble, will be referred to as {\it
extended} $N$-centered ensemble formalism. 

\subsection{Combining TGOK with $N$-centered ensembles}

By analogy with regular $N$-centered ensemble DFT, where the ensemble
weights assigned to the $(N-1)$- and $(N+1)$-electron ground
states are allowed to vary independently~\cite{senjean2018unified}, we
propose to combine TGOK and $N$-centered ($N$c) ensembles as follows, 
\be\label{eq:eNc_ens_matrix_op}
\hat{\Gamma}^{\bxi}\overset{{\rm TGOK}+N{\rm c}}{=}
\qty(1 - \sum_{\nu>0}\dfrac{N_\nu}{N}\xinu)\dyad*{{\Psi}_0}
+\sum_{\nu>0}\xinu\dyad*{{\Psi}_{\nu}},
\ee
where $\Psi_0\equiv {\Psi}_0^N$ is the reference $N$-electron ground
state to which all possible excitation processes (neutral and charged, including multiple-electron excitations) can be applied. In other words,
the integer number of electrons $N_\nu=\int d\br\,n_{\Psi_\nu}(\br)$ that is described by the excited-state
wave function $\Psi_\nu$ ($\nu>0$) is not necessarily equal to the
(integer) number of electrons in the ground state, 
\be\label{notation_central_nbr_electrons}
N_0=N,
\ee
that is referred to as the central number of electrons. 
However, by construction, the ensemble density still
integrates to $N$, like in TGOK-DFT [see
Eqs.~(\ref{eq:ens_dens_integrates_to_N}) and
(\ref{eq:ens_dens_from_Gamma_hat})]. 

As further discussed in Sec.~\ref{sec:DDs_revisited}, the fact that the net number of
electrons in the ensemble does {\it not} vary with the ensemble weights,
unlike in the traditional PPLB-DFT of fractional electron
numbers~\cite{perdew1982density}, plays a central role in the exact
description of derivative discontinuities in terms of ensemble
weight derivatives~\cite{PRA21_Hodgson_exact_Nc-eDFT_1D,Cernatic2022}.
More precisely, the key feature of $N$-centered ensemble DFT is the fact that
ensemble weights can vary while holding the ensemble density
fixed~\cite{senjean2018unified}.
In PPLB-DFT, the Hxc functional has no weight dependence (\ie, it is
solely a functional of the density) simply because
the ensemble weight is itself a functional of the
density~\cite{Cernatic2022} which can be defined, for example, as the deviation
of the (fractional) electron
number from its floor value.
Hence, in PPLB-DFT, any variation in weight automatically
induces a change in density.       

Turning back to the extended $N$-centered ensemble of
Eq.~(\ref{eq:eNc_ens_matrix_op}), we point out that
the ensemble weights assigned to
the ground and excited states are positive,
\begin{subequations}
\label{eq:Nc_convexity_constraints}
\begin{align}
\label{eq:eNc_weight_GS_general}
\xi_0&=1-\sum_{\nu>0}\dfrac{N_\nu}{N}\xinu\geq 0,
\\
\xi_\nu&\geq 0,\quad \forall\nu>0,
\end{align}
\end{subequations}
and, most importantly, unlike in conventional TGOK or PPLB ensembles, they do {\it not} sum up
to 1, in general: 
\be
\sum_{\nu\geq 0}\xi_\nu=
1+\sum_{\nu>0}\dfrac{\left(N-N_\nu\right)}{N}\xinu.
\ee
In addition, within each subensemble containing all the states with the same number $N\pm p$ ($p=0,1,2,\ldots,N$) of electrons, we impose the following weight ordering constraints
[see Eq.~(\ref{TGOK_weights_ordering})],
\be\label{eq:Nc_TGOK_constraints}
\xinu\overset{N_{\nu}=N_{\nu+1}=N\pm p}{\ge}\xi_{\nu+1},
\ee
in order to be able to exploit, if necessary, the TGOK variational principle within each $(N\pm
p)$-electron
sector of the Fock space. This simply ensures that the fully extended
$N$-centered ensemble of interest can be determined variationally, thus allowing
for an in-principle-exact density-functional description of its energy. Note that, in the present work, the constraint of
Eq.~(\ref{eq:Nc_TGOK_constraints}) will be used only for $N$-electron
states (\ie, $p=0$) while the $(N-1)$ sector will be reduced to
ground states. The $(N+1)$ sector will
not be
used. 
These choices are motivated by the problem raised
in Sec.~\ref{sec:problematic} and are by no means a limitation of the
present ensemble formalism, which is very general.\\

Let us finally turn to the extended $N$-centered ensemble energy,
\be\label{eq:eNc-ens_ener_def}
E^{\bxi}\overset{{\rm TGOK}+N{\rm c}}{=}
\qty(1 - \sum_{\nu>0}\dfrac{N_\nu}{N}\xinu)E_0
+\sum_{\nu>0}\xinu E_\nu,
\ee
where $E_0=E_0^N$ is the reference $N$-electron ground-state energy and
$E_\nu$ is an $N_\nu$-electron eigenvalue of the electronic Hamiltonian, \ie,
\be
\hat{H}\ket{\Psi_\nu}=E_\nu\ket{\Psi_\nu}, \quad \nu\geq 0,
\ee
with $N_\nu\in
\left\{N,N\pm1,N\pm2,\ldots\right\}$. Like in
TGOK-DFT~\cite{deur2019ground}, the linearity of the ensemble
energy in the ensemble weights,
\be\label{eq:eNc_weights_notation_bxi}
\bxi
\overset{{\rm TGOK}+N{\rm c}}{\equiv}
 \left\{\xi_\nu\right\}_{\nu>0},
\ee 
allows for a straightforward extraction of individual energy levels
(and, therefore, of the excitation energies) through first-order
differentiations. Indeed, since both ground-state ($\nu=0$) [see Eq.~(\ref{notation_central_nbr_electrons})] and excited-state
($\nu>0$) energies can be expressed as follows,   
\be\label{eq:E_nu_exp}
E_{\nu}\underset{\nu\ge0}{=}\dfrac{N_\nu}{N}E_0+\sum_{\lambda>0}\delta_{\lambda\nu}\left(E_\lambda-\dfrac{N_\lambda}{N}E_0\right),
\ee
where, according to Eq.~(\ref{eq:eNc-ens_ener_def}),
\be\label{eq:E_lambda_minus_E0}
\dfrac{\partial E^{\bxi}}{\partial
\xi_\lambda}=E_\lambda-\dfrac{N_\lambda}{N}E_0,
\ee
and
\be\label{eq:E0_from_ens_ener}
E_0=E^{\bxi}-\sum_{\lambda>0}\xi_\lambda\dfrac{\partial E^{\bxi}}{\partial
\xi_\lambda},
\ee
we immediately obtain the following compact expression in terms of the
ensemble energy (and its first-order derivatives), 
\be\label{eq:eNc-ens_indiv-energies}
E_{\nu}\underset{\nu\ge0}{=}\dfrac{N_\nu}{N}E^{\bxi}+
\sum_{\lambda>0}\qty( \delta_{\lambda\nu}-\dfrac{N_\nu}{N}\xilambda)\dfrac{\partial E^{\bxi}}{\partial
\xi_\lambda}.
\ee
Note that Eq.~(\ref{eq:eNc-ens_indiv-energies}) generalizes 
expressions that have been derived previously and separately in regular
TGOK~\cite{deur2019ground} (for neutral excited states) and
$N$-centered~\cite{senjean2018unified,Cernatic2022} (for charged excited
states)
ensemble theories. As shown in the following section, once a KS
density-functional description of the extended $N$-centered ensemble
energy $E^{\bxi}$ is established, it becomes possible
to relate formally any charged or neutral KS excitation energy to the true
physical one.

\subsection{Density-functionalization of the approach}

According to both regular (ground-state) Rayleigh--Ritz and
TGOK~\cite{JPC79_Theophilou_equi-ensembles,gross1988rayleigh}
variational principles, the extended $N$-centered ensemble
energy of Eq.~(\ref{eq:eNc-ens_ener_def}) can be determined
variationally, thus allowing for its in-principle-exact ensemble
density-functional description. The exact same formalism as in TGOK-DFT
can actually be used [see
Eqs.~(\ref{eq:variational_ens_ener_KS_scheme})--(\ref{eq:total_KS_ener_from_KS_orb_ener})]. The only difference is that we are now allowed to consider occupations
in the KS wave functions [see Eq.~(\ref{eq:total_KS_ener_from_KS_orb_ener})] that do not necessarily sum up to $N$: 
\be\label{eq:sum_occu_KS_wfs}
\sum_i n_\nu^i=N_\nu.
\ee

By rewriting the ensemble energy as follows [see
Eqs.~(\ref{eq:variational_ens_ener_KS_scheme}),
(\ref{eq:ens_KS_eqs_many-body_KS_states}), and (\ref{eq:ens_dens_from_KS_many_body_states})]
\be\label{eq:ens_ener_in_terms_KS_energies}
E^{\bxi}=\sum_{\nu\geq 0}\xi_\nu\mathcal{E}^{\bxi}_\nu+E_{\rm
Hxc}^{\bxi}[n^{\bxi}]-\int d\br\,v_{\rm
Hxc}^{\bxi}(\br)n^{\bxi}(\br),
\ee
and applying the Hellmann-Feynman theorem to its variational
expression in Eq.~(\ref{eq:variational_ens_ener_KS_scheme}), 
which leads to 
\be\label{eq:deriv_ens_ener_HF_theo}
\dfrac{\partial E^{\bxi}}{\partial
\xi_\lambda}
=\mathcal{E}^{\bxi}_\lambda-\dfrac{N_\lambda}{N}\mathcal{E}^{\bxi}_0
+\left.\dxilambda{E_{\rm Hxc}^{\bxi}[n]}\right|_{n=n^{\bxi}},
\ee
we deduce from Eq.~(\ref{eq:eNc-ens_indiv-energies}) the
following exact expression for any individual energy level included
into the ensemble
\be\label{eq:eNc_indiv-energies-from-KS}
\begin{split}
E_{\nu}\underset{\nu\geq0}{=}\mathcal{E}^{\bxi}_\nu&+\dfrac{N_\nu}{N}\left(E_{\rm Hxc}^{\bxi}[n^{\bxi}]-\int d\br\, v_{\rm Hxc}^{\bxi}(\br)n^{\bxi}(\br)\right)
\\
&
+\sum_{\lambda>0}\qty(
\delta_{\lambda\nu}-\dfrac{N_\nu}{N}\xilambda)\left.\dxilambda{E_{\rm
Hxc}^{\bxi}[n]}\right|_{n=n^{\bxi}},
\end{split}
\ee
where, in the second term on the right-hand side, we recognize the
analog for ensembles of the Levy--Zahariev shift in
potential~\cite{levy2014ground,deur2019ground,senjean2018unified}.
\rev{Note in passing that a similar shift was actually recovered in an
earlier work by Kraisler and Kronik~\cite{kraisler2013piecewise} where
the authors described the piecewise-linear behavior of the ground-state
energy within ensemble DFT for fractional electron numbers.} At this point we should make an important observation that will be
exploited later on, namely that the above expression is invariant
under constant
shifts in the ensemble Hxc potential, $v_{\rm Hxc}^{\bxi}(\br)\rightarrow
v_{\rm Hxc}^{\bxi}(\br)+c$, since [see
Eqs.~(\ref{eq:ens_dens_integrates_to_N}), (\ref{eq:total_KS_ener_from_KS_orb_ener}), and (\ref{eq:sum_occu_KS_wfs})]
\begin{multline}\label{eq:invariance_constant_shifts_NceDFT}
\mathcal{E}^{\bxi}_\nu-\dfrac{N_\nu}{N}\int d\br\, v_{\rm Hxc}^{\bxi}(\br)n^{\bxi}(\br)
\\
= \left(\mathcal{E}^{\bxi}_\nu+N_\nu c\right)-\dfrac{N_\nu}{N}\int
d\br\, \left(v_{\rm Hxc}^{\bxi}(\br)+c\right)n^{\bxi}(\br),
\end{multline}
even though the ensemble may contain states that describe different
numbers of electrons. This major difference between (extended or not) 
$N$-centered ensemble DFT and the conventional DFT for fractional
electron numbers originates from the fact that, in the former theory,
the number of electrons is artificially held constant and equal to the
integer $N$ [see Eq.~(\ref{eq:ens_dens_integrates_to_N})].
\rev{This flexibility allows us to recover, in the $N$-electron
ground-state limit of the theory [in this case, the ensemble weights are set to $\bxi=0$ in
Eq.~(\ref{eq:eNc_indiv-energies-from-KS})], the Hxc potential of regular $N$-electron ground-state
KS-DFT, which is also uniquely defined up to a constant. This point is more thoroughly 
discussed for the fundamental gap problem in
Refs.~\citenum{PRA21_Hodgson_exact_Nc-eDFT_1D} and
\citenum{Cernatic2022}. It is also important to note that setting $\bxi=0$ in
Eq.~(\ref{eq:eNc_indiv-energies-from-KS}) consists in fact in taking the
limit $\bxi\rightarrow 0^+$ of extended $N$-centered ensemble KS-DFT,
because infinitesimal deviations from the zero weight limit $\bxi=0$ [which
does correspond to regular $N$-electron ground-state
KS-DFT, according to Eqs.~(\ref{eq:variational_ens_ener_KS_scheme}),
(\ref{eq:eNc-ens_ener_def}), and (\ref{eq:eNc_weights_notation_bxi})]
are needed to evaluate weight derivatives of the ensemble Hxc density functional
[second line of Eq.~(\ref{eq:eNc_indiv-energies-from-KS})]. This is how
any (charged or neutral) excited-state energy level can be described, in
principle exactly, through the extended $N$-centered ensemble formalism (\ie, by
considering the $\bxi\rightarrow 0^+$ limit), on the basis of regular
DFT ($\bxi=0$), which provides a variational description of the $N$-electron
ground-state energy only.}\\ 

We finally conclude from Eq.~(\ref{eq:eNc_indiv-energies-from-KS}) that
the energy associated to
any (charged or neutral) excitation $\nu\rightarrow\kappa
$ can be expressed exactly in terms of its KS analog as follows
\be\label{eq:any_XE_in_terms_KS_XE}
\begin{split}
&E_\kappa-E_\nu=\mathcal{E}^{\bxi}_\kappa-\mathcal{E}^{\bxi}_\nu
\\
&
+\dfrac{\left(N_\kappa-N_\nu\right)}{N}\left(E_{\rm Hxc}^{\bxi}[n^{\bxi}]-\int d\br\, v_{\rm Hxc}^{\bxi}(\br)n^{\bxi}(\br)\right)
\\
&
+
\sum_{\lambda>0}\qty(
\delta_{\lambda\kappa}-\delta_{\lambda\nu}-\dfrac{\left(N_\kappa-N_\nu\right)}{N}\xilambda)\left.\dxilambda{E_{\rm
Hxc}^{\bxi}[n]}\right|_{n=n^{\bxi}}.
\end{split}
\ee
Equation (\ref{eq:any_XE_in_terms_KS_XE}) is our first key result. It
generalizes the neutral excitation
energy expression of TGOK-DFT~\cite{gross1988density,deur2019ground}
that was recalled in Eq.~(\ref{eq:excitation_energy_from_TGOK-DFT}).

\section{Revisiting density-functional derivative discontinuities induced by neutral excitations}\label{sec:DDs_revisited}

In order to achieve an exactification of the KS orbital energies along the
lines of Sec.~\ref{sec:problematic}, we apply the general formalism of
Sec.~\ref{sec:eNc_ensemble_DFT} to a particular type of extended
$N$-centered ensemble which includes only the neutrally-excited $N$-electron
states, with weights $\left\{\xi_\lambda^N\right\}_{\lambda>0}$, {\it and} the ionized $(N-1)$-electron
ground state, with weight $\xi_-:=\xi^{N-1}_0$. Therefore, in this
special case, the collection of ensemble weights reduces to 
\be\label{eq:weight_notation}
\bxi\equiv\left(\left\{\xi^N_\lambda\right\}_{\lambda>0},\xi_-\right),
\ee
\rev{and Eq.~(\ref{eq:any_XE_in_terms_KS_XE}), where the summation in
$\lambda>0$ (last term on the right-hand side) runs over both the
neutral excited states $\Psi^N_\lambda$ (to which the weights $\xi_\lambda^N$ are
assigned) and the ground ionized state $\Psi^{N-1}_0$ simply labelled as
``$\lambda=-$'' (to which the
weight $\xi_-$ is assigned), becomes
\be
\begin{split}
&E_\kappa-E_\nu=\mathcal{E}^{\bxi}_\kappa-\mathcal{E}^{\bxi}_\nu
\\
&
+\dfrac{\left(N_\kappa-N_\nu\right)}{N}\left(E_{\rm Hxc}^{\bxi}[n^{\bxi}]-\int d\br\, v_{\rm Hxc}^{\bxi}(\br)n^{\bxi}(\br)\right)
\\
&
+
\sum_{\lambda>0}\qty(
\delta_{\lambda\kappa}-\delta_{\lambda\nu}-\dfrac{\left(N_\kappa-N_\nu\right)}{N}\xi^N_\lambda)\left.\dfrac{\partial E_{\rm
Hxc}^{\bxi}[n]}{\partial\xi^N_\lambda}\right|_{n=n^{\bxi}}
\\
&
+
\qty(
\delta_{-\kappa}-\delta_{-\nu}-\dfrac{\left(N_\kappa-N_\nu\right)}{N}\xi_-)\left.\dfrac{\partial E_{\rm
Hxc}^{\bxi}[n]}{\partial \xi_-}\right|_{n=n^{\bxi}}
.
\end{split}
\ee
If, in the above expression, we choose for $\kappa$ the ground ionized
state (\ie, $\delta_{\lambda\kappa}=0$ in the third line,
$\delta_{-\kappa}=1$ in the fourth line, $N_\kappa=N-1$,
$E_\kappa=E^{N-1}_0$, and
$\mathcal{E}^{\bxi}_\kappa=\sum^{N-1}_{i=1}\varepsilon^{\bxi}_i$), and
for $\nu\geq 0$, any ground or excited $N$-electron
state (\ie, $\delta_{-\nu}=0$ in the fourth line, $N_\nu=N$, and
$E_\nu=E^N_\nu$), then we can express any ground- or
excited-state IP as $I^N_\nu=E_\kappa-E_\nu$ [see
Eq.~(\ref{eq:IP_N-electron_nu-state})]. This gives the following more
explicit exact expression,
\be
\begin{split}
I^N_\nu\underset{\nu\geq0}{=}&\sum^{N-1}_{i=1}\varepsilon^{\bxi}_i-\mathcal{E}^{\bxi}_\nu
\\
&
-\dfrac{1}{N}\left(E_{\rm Hxc}^{\bxi}[n^{\bxi}]-\int d\br\, v_{\rm Hxc}^{\bxi}(\br)n^{\bxi}(\br)\right)
\\
&
+
\sum_{\lambda>0}\qty(
-\delta_{\lambda\nu}+\dfrac{\xi^N_\lambda}{N})\left.\dfrac{\partial E_{\rm
Hxc}^{\bxi}[n]}{\partial\xi^N_\lambda}\right|_{n=n^{\bxi}}
\\
&
+
\qty(
1+\dfrac{\xi_-}{N})\left.\dfrac{\partial E_{\rm
Hxc}^{\bxi}[n]}{\partial \xi_-}\right|_{n=n^{\bxi}}
,
\end{split}
\ee
thus leading to 
}
\be\label{eq:general_exp_I_nu_eNc_eDFT}
\begin{split}
I^N_\nu
\underset{\nu\geq0}{=}\rev{-\varepsilon^{\bxi}_{i(\nu)}}
& -\dfrac{1}{N}\left(E_{\rm Hxc}^{\bxi}[n^{\bxi}] -\int d\br\, v_{\rm Hxc}^{\bxi}(\br)n^{\bxi}(\br)\right)
\\
&
+\left(1+\dfrac{\xi_-}{N}\right)\left.\dfrac{\partial E_{\rm
Hxc}^{\bxi}[n]}{\partial \xi_-}\right|_{n=n^{\bxi}}
\\
&
+
\sum_{\lambda>0}
\qty(
\dfrac{\xi^N_\lambda}{N}
-\delta_{\lambda\nu}
)\left.\dfrac{\partial E_{\rm
Hxc}^{\bxi}[n]}{\partial \xi^N_\lambda}\right|_{n=n^{\bxi}},
\end{split}
\ee
\rev{where, along the lines of Sec.~\ref{sec:problematic}, we assumed,
for simplicity and clarity, that within the KS ensemble, the neutral excited states
$\nu>0$ of interest are described by a single-electron excitation from
the HOMO $\varphi^{\bxi}_N=:\varphi^{\bxi}_{i(0)}$} to \rev{a higher-energy orbital that we
denote $\varphi^{\bxi}_{i(\nu)}$ (where the index $i(\nu)>N$ of the
acceptor orbital depends on the excited state of interest
$\nu>0$)}, \ie, its energy reads as follows in terms of the KS orbital energies,
$\mathcal{E}^{\bxi}_\nu=\sum^{N-1}_{i=1}\varepsilon^{\bxi}_i+\rev{\varepsilon^{\bxi}_{i(\nu)}}$,
the HOMO energy being
\rev{$\varepsilon^{\bxi}_N=:\varepsilon^{\bxi}_{i(0)}$}. 

The description of \rev{both single excitations from lower-energy orbitals and} double
excitations~\cite{sagredo2018can,Gould2021_Double,marut2020weight} within the present
formalism, as well as how
excitations in the true interacting system are connected to those
occurring in the
noninteracting KS ensemble, is discussed in a separate work~\cite{cernatic2024extended_doubles}. Still, it is important to stress that, whatever
the nature of the neutral excitation $\nu>0$ under study, Eq.~(\ref{eq:any_XE_in_terms_KS_XE}),
from which excited-state IPs can be evaluated,
remains exact. The possible mixture of single and double (or even higher) excitations in
the true interacting states, which may not appear explicitly in the KS states, will
be recovered, energy-wise, {\it via} the ensemble Hxc functional and its weight
derivatives [see the second and third lines of
Eq.~(\ref{eq:any_XE_in_terms_KS_XE})].\\   

As pointed out previously [see
Eq.~(\ref{eq:invariance_constant_shifts_NceDFT})], the IP expression of
Eq.~(\ref{eq:general_exp_I_nu_eNc_eDFT}) is invariant under any
constant shift in the ensemble Hxc potential. Therefore, we can always
adjust the latter potential in order to exactify Koopmans' theorem
for a given ground or excited $N$-electron state $\nu$ and a given
choice of ensemble weight values $\bxi$:
\be\label{eq:exact_Koopmans_theo_eNc-eDFT}
\begin{split}
I^N_\nu&\rev{\underset{\nu\geq 0}{=}-\varepsilon^{\bxi}_{i(\nu)}}
\\
&\Updownarrow 
\\
\int d\br\, v_{\rm Hxc}^{\bxi}(\br)n^{\bxi}(\br)&=E_{\rm
Hxc}^{\bxi}[n^{\bxi}]
\\
&
-\left(N+\xi_-\right)\left.\dfrac{\partial E_{\rm
Hxc}^{\bxi}[n]}{\partial \xi_-}\right|_{n=n^{\bxi}}
\\
&
+\sum_{\lambda>0}
\qty(
N\delta_{\lambda\nu}
-\xi^N_\lambda
)\left.\dfrac{\partial E_{\rm
Hxc}^{\bxi}[n]}{\partial \xi^N_\lambda}\right|_{n=n^{\bxi}}.
\end{split}
\ee    
Equation (\ref{eq:exact_Koopmans_theo_eNc-eDFT}), which is the second key
result of this work, uniquely defines (not up to
a constant anymore) the Hxc potential. Interestingly, unlike in
traditional DFT approaches to electronic excitations~\cite{gould2019asymptotic}, this alternative
and explicit adjustment procedure of the Hxc potential does {\it not} rely on the asymptotic
behavior of the density (see Refs.~\onlinecite{PRA21_Hodgson_exact_Nc-eDFT_1D,Cernatic2022} for a comparison of the
two formalisms in the ground state), which means that it is not only
applicable to {\it ab initio} molecular systems but it should
also be transferable to finite-size lattice models or extended systems,
in principle.
\\  

Finally, in order to revisit the concept of derivative discontinuity
within the present ensemble density-functional formalism, let us have
a closer look at the ionization of both the ground $N$-electron state
(first scenario) and a specific
$\nu$th neutral excited state (second scenario), separately. For that
purpose, we should first realize that, in order to reach the
latter state $\nu$ variationally, we only need to include into the ensemble the neutral states that are lower in
energy than $\nu$, \ie, we can set to zero the ensemble weights
$\xi^N_\lambda$ corresponding to $\lambda>\nu$. In this case, the collection of ensemble weights further
reduces to 
\be
\bxi\equiv\left(\bxi^N_\nu,0,0,\ldots,0,\xi_-\right),
\ee
where (note the bold font) 
\be\label{eq:weight_notation_2}
\bxi^N_\nu\equiv \left(\xi^N_1,\xi^N_2,\ldots,\xi^N_\nu\right)
\ee
is a shorthand notation for the $\nu$ non-zero and monotonically
decreasing (TGOK) ensemble weights. In the first scenario, we adjust the Hxc
potential such that Koopmans' theorem is fulfilled for the ground state
($\nu=0$). As neutral excited states are not involved {\it at all} in this case,
the ensemble can be reduced to a regular $N$-centered
ensemble~\cite{senjean2018unified} where the TGOK ensemble weights are
strictly set to zero:  
\be\label{eq:scenario1}
\bxi_1\equiv\left(\bxi^N_\nu={\bm 0},0,0,\ldots,0,\xi_->0\right).
\ee
On the other hand, in the second scenario (ionization of the $\nu$th
excited state), assigning an infinitesimal weight to the ionized ground
state, in addition to strictly positive and monotonically decreasing
weights in $\bxi^N_\nu$, is sufficient, {\ie},
\be\label{eq:scenario2}
\bxi_2\equiv\left(\bxi^N_\nu>0,0,0,\ldots,0,\xi_-\rightarrow 0^+\right),
\ee
so that the ensemble weight derivative $\partial E_{\rm
Hxc}^{\bxi}[n]/\partial \xi_-$ in
Eq.~(\ref{eq:exact_Koopmans_theo_eNc-eDFT}) can still be evaluated, like
in the first scenario. As a
consequence of Eq.~(\ref{eq:exact_Koopmans_theo_eNc-eDFT}), we finally
reach from Eq.~(\ref{eq:neutral_XE_from_IPs}), and without ever invoking fractional electron numbers nor
referring to the
asymptotic behavior of the ensemble density, the desired exactification
of the KS orbital energies, 
\be\label{eq:Exact_Omega_from_KS_orb_energies}
E^N_\nu-E_0^N\rev{\underset{\nu>0}{=}}\rev{\varepsilon^{\bxi_2}_{i(\nu)}}-\varepsilon^{\bxi_1}_{N},
\ee
\rev{where $i(\nu)>N$}, with a clear and explicit construction of the
corresponding Hxc potentials. 

Equation (\ref{eq:Exact_Omega_from_KS_orb_energies}) generalizes Levy's exact formula for the optical
gap~\cite{levy1995excitation}. It also offers a different perspective on
its more recent extension to higher neutral single-electron
excitations~\cite{gould2022single}.
Most importantly, the first and second scenarios have a connection point which is reached
when $\xi_-\rightarrow 0^+$ and $\bxi^N_\nu\rightarrow {\bm 0}^+$,
respectively, and which corresponds to the regular $N$-electron
ground-state formulation of DFT (the density equals $n_{\Psi^N_0}$ in
this case). If we use the shorthand notation 
\be
\bxi^N_\nu\underset{\rm notation}{\equiv}\left(\bxi^N_\nu,0,0,\ldots,0,\xi_-\rightarrow
0^+\right),
\ee
{then equating, at this connection point,  the Hxc potentials
associated with these two scenarios via Eq.~(\ref{eq:exact_Koopmans_theo_eNc-eDFT}) leads to our third key result:}
\be\label{eq:eq:Exact_Omega_from_KS_orb_energies_zero_weight_limit}
E^N_\nu-E_0^N=\rev{\varepsilon^{\bxi^N_\nu\rightarrow {\bm
0}^+}_{i(\nu)}}-\varepsilon^{\bxi^N_\nu={\bm 0}}_{N},
\ee
where
\be\label{eq:HxcDD_from_NceDFT_final_result}
\begin{split}
&\int\dfrac{d\mathbf{r}}{N}\left(v_{\rm Hxc}^{\bxi^N_\nu\rightarrow {\bm
0}^+}(\mathbf{r})-v_{\rm Hxc}^{\bxi^N_\nu={\bm 0}}(\mathbf{r})\right)
n_{\Psi_0^N}(\mathbf{r})
\\
&=\left.\dfrac{\partial E_{\rm
Hxc}^{\bxi^N_\nu}[n_{\Psi_0^N}]}{\partial\xi^N_\nu}\right|_{\bxi^N_\nu={\bm 0}}.
\end{split}
\ee
If we use the following decomposition of the ensemble Hxc energy in terms of the regular weight-independent Hartree functional and the
weight-dependent xc functional,
\be\label{eq:GSH_plus_ens_xc}
E_{\rm Hxc}^{\bxi}[n]=E_{\rm H}[n]+E_{\rm xc}^{\bxi}[n],
\ee
which is formally convenient but problematic in practice, because
of the ghost interaction errors it may
induce~\cite{ensemble_ghost_interaction,Cernatic2022}, 
then all Hartree terms can be removed
from Eq.~(\ref{eq:HxcDD_from_NceDFT_final_result}). 
Thus we recover, from a completely different ($N$-centered) ensemble perspective, a
feature that was originally highlighted by
Levy~\cite{levy1995excitation}, namely that the exactification of
neutral KS excitation energies is conditioned by the appearance of a
derivative discontinuity in the xc potential, once the excitation of
interest has been incorporated into the ensemble. Moreover, as readily
seen from Eq.~(\ref{eq:HxcDD_from_NceDFT_final_result}),
that derivative discontinuity
matches the weight derivative \rev{(taken at $\bxi^N_\nu={\bm
0}$) of the ensemble Hxc functional, which, on the other hand, corresponds to the
deviation in excitation energy between the physical and KS systems [see
Eq.~(\ref{eq:excitation_energy_from_TGOK-DFT})]. The equivalence of the
two concepts in this special limit,
already mentioned at the end of Sec.~\ref{sec:review}, is recovered from Eqs.~(\ref{eq:eq:Exact_Omega_from_KS_orb_energies_zero_weight_limit}) and (\ref{eq:HxcDD_from_NceDFT_final_result}).}

Equation (\ref{eq:eq:Exact_Omega_from_KS_orb_energies_zero_weight_limit})
echoes Eq.~(1) of Ref.~\onlinecite{gould2022single} taken in the
so-called
``$w\rightarrow 0 ^+$'' limit, which is the foundation of perturbative ensemble
DFT [pEDFT]~\cite{gould2022single}. It explains how exact neutral excitation energies
can be evaluated from extended $N$-centered ensemble limits towards regular $N$-electron ground-state DFT.
From a practical point of view, Eq.~(\ref{eq:eq:Exact_Omega_from_KS_orb_energies_zero_weight_limit})
offers an alternative approach, in the evaluation of gaps, to the still challenging design of weight-dependent
density-functional approximations~\cite{loos2020weightdependent}. While the latter
would in principle allow for a straightforward
evaluation of the former from a single ensemble DFT
calculation (see
Eq.~(\ref{eq:excitation_energy_from_TGOK-DFT}) and
Ref.~\onlinecite{deur2019ground}), two different limits towards regular
DFT need to be considered instead. In both of them, the $N$-centered ensemble
weight $\xi_-$ infinitesimally deviates from zero. Revisiting pEDFT in the
present context deserves further attention. This is left for future
work.   

\section{Application to the Hubbard dimer}\label{sec:Application_Hdim}

The present section is a proof of concept, illustrating how derivative
discontinuities emerge for neutral electronic excitations
in the context of extended $N$-centered (e$N$-centered) ensemble DFT.
For this purpose, the e$N$-centered ensemble
formalism is applied to both symmetric and asymmetric Hubbard dimer
models~\cite{carrascal2015hubbard}. 

\subsection{Key features of the model}\label{subsec:Hdim_Key_features}

The Hubbard dimer has become in recent years a model of
choice for assessing density-functional approximations in various
contexts but also for exploring new concepts~\cite{carrascal2015hubbard,li2018density,deur2018exploring,
sagredo2018can,carrascal2018linear,smith2016exact,deur2019ground,
Cernatic2022,Giarrusso2023_Exact,Ullrich2018_Density,scott2023exact,Sobrino2023_What,Liebert2023_Refining}. 
In the (second-quantized) Hamiltonian
$\hat{\mathcal{H}}=\hat{\mathcal{T}}+\hat{\mathcal{U}}+\hat{V}_{\rm
ext}$ of the Hubbard dimer model, the kinetic energy, electron repulsion, and local
(external) potential operators are simplified as follows,
\begin{subequations}\label{eq:Hamiltonian_Hdim}
\begin{align}
\hat{\mathcal{T}}&=-t\sum_{\sigma = \uparrow,\downarrow}
(\hat{c}^{\dagger}_{0\sigma}\hat{c}_{1\sigma} + \hat{c}^{\dagger}_{1\sigma}\hat{c}_{0\sigma}),
\label{eq:Hamiltonian_Hdim_hopping}
\\
\hat{\mathcal{U}}&=U\sum_{i=0}^{1}\hat{n}_{i\uparrow}\hat{n}_{i\downarrow},
\label{eq:Hamiltonian_Hdim_U}
\\
\hat{V}_{\rm ext}&=\frac{\Delta v_{\rm ext}}{2}(\hat{n}_{1} - \hat{n}_{0}),
\label{eq:Hamiltonian_Hdim_vext}
\end{align}
\end{subequations}
where $i \in \qty{0,1}$ labels the two atomic sites,  $\hat{n}_{i\sigma} = \hat{c}^{\dagger}_{i\sigma}\hat{c}_{i\sigma}$
is the spin-site occupation operator, and $\hat{n}_{i} = \sum_{\sigma =
\uparrow,\downarrow}\hat{n}_{i\sigma}$ plays the role of
the density operator. While the difference in external potential $\Delta
v_{\rm ext}$ controls the asymmetry of the model, the ratio $U/t$ of the
on-site electronic repulsion parameter $U$ to the hopping parameter $t$
can be used to tune electron correlation effects in the model. In the
following, the central number of electrons in the e$N$-centered
ensemble is set to $N=2$.  With this choice, the ensemble electronic density
$\left\{n_i=\langle\hat{n}_i\rangle\right\}_{i=0,1}$ reduces to a single
variable $n$ with site
occupations equal to $n_{0}=n$ and $n_{1} = 2-n$ [see
Eq.~(\ref{eq:ens_dens_integrates_to_N})]. Note that $n=1$ in the
symmetric dimer.

As for the construction of the e$N$-centered ensemble, we consider the singlet $N$-electron ground state,
which also determines the choice for the ($N-1$)-electron ground state, that is, the one-electron ground state of
the noninteracting ($U=0$) Hamiltonian. We also include into the
ensemble the lowest singlet $N$-electron (neutral) excited state which, in the
noninteracting limit, is described by a single electron (HOMO to LUMO)
excitation. The collection $\bxi\equiv(\xi,\xi_-)$ of e$N$-centered
ensemble weights consists of the weight $\xi$ assigned to the neutral 
(two-electron) excited state and the weight $\xi_-$ that is assigned to the
one-electron ground state.
\rev{According to Eq.~(\ref{eq:eNc_weight_GS_general}), the e$N$-centered
ensemble weight assigned to the reference two-electron ground state
($N=2$ in this case) reads
\be\label{eq:xi0_HD_general_exp}
\xi_0\equiv \xi_0(\bxi)=1-\dfrac{(N-1)\xi_-}{N}-\xi=1-\dfrac{\xi_-}{2}-\xi.
\ee
}
Note that the ensemble weight constraints of Eqs.~\eqref{eq:Nc_convexity_constraints} and
\eqref{eq:Nc_TGOK_constraints} read
\rev{
\begin{subequations}
\begin{align}
\label{eq:xi0_HD}
 &\xi_0=1-\dfrac{\xi_-}{2}-\xi\geq 0,
\\
\label{eq:xi_minus_positive_HD}
&\xi_-\geq 0,
\end{align}
\end{subequations}
and
\be\label{eq:TGOK_weight_order_HD}
\xi_0\geq \xi\geq 0,
\ee
respectively,
or, equivalently [according to Eqs.~(\ref{eq:xi0_HD}) and (\ref{eq:TGOK_weight_order_HD})],
\be
1-\dfrac{\xi_-}{2}-\xi\geq \xi
\ee
and [according to Eqs.~(\ref{eq:xi0_HD}) and (\ref{eq:xi_minus_positive_HD})]
\be
0\leq \xi_-\leq 2\left(1-\xi\right),
\ee
thus leading to
 $0 \leq \xi \leq \frac{1}{2} -
({\xi_-}/{4})$
for $\xi_-$ varying in $0 \leq \xi_- \leq 2$. The e$N$-centered
KS ensemble is described by two fractionally occupied orbitals that we simply
refer to as HOMO and LUMO in the following, as we use the two-electron ground state as
reference in the ensemble. Their expansions in the site basis as well as
their energies can be obtained by diagonalizing
the following one-electron KS Hamiltonian matrix [see Eq.~(\ref{eq:Hamiltonian_Hdim})],
\be\label{eq:one-elec_KS_hamil_matrix}
\left[\hat{h}^{\rm KS}\left(\Delta v\right)\right]=
\left[
\begin{matrix}
-\dfrac{\Delta v}{2} & -t
\\
-t & +\dfrac{\Delta v}{2}
\end{matrix}
\right]
.
\ee
In Eq.~(\ref{eq:one-elec_KS_hamil_matrix}), $\Delta v$ corresponds to the ensemble density-functional KS
potential difference whose exact expression is given later in
Eq.~(\ref{eq:eNc_KS_dens_fun_pot_HD}) [see the
proof that follows Eq.~(\ref{eq:TS-LFTransform_Hdim}) in
Appendix~\ref{sec:appendix_exact_functionals}]. According to
Eq.~(\ref{eq:xi0_HD_general_exp}), the fractional occupations
$n^{\bxi}_{i}$ ($i=$ H or L) of the
HOMO and LUMO are controlled by the ensemble weights $\bxi$ as follows 
[see Eq.~(\ref{eq:ens_dens_from_KS_many_body_states})],
\be
\begin{split}
\label{eq:occ_homo_HD}
n^{\bxi}_{\rm H}&=\xi_0(\bxi)\times 2+\xi_-\times 1+\xi\times 1
\\
&=2-\xi,
\end{split}
\ee
and
\be
\label{eq:occ_lumo_HD}
\begin{split}
n^{\bxi}_{\rm L}&=\xi_0(\bxi)\times 0+\xi_-\times 0+\xi\times 1
\\
&=\xi,
\end{split}
\ee
respectively.} Note finally that, for a given
choice of $\bxi$, we can arbitrarily shift the external potential (or the KS
potential) introduced in Eq.~(\ref{eq:Hamiltonian_Hdim}) by a constant, {\ie}
\be\label{eq:shifting_pot_HD}
\pm\dfrac{\Delta v}{2} \rightarrow \pm\dfrac{\Delta v}{2}-\mu,
\ee
without
modifying the e$N$-centered ensemble under study. As further discussed in the following, the
adjustment of the latter constant in the KS potential will be essential
for satisfying exact 
Koopmans' theorems and, consequently, observing Hxc derivative
discontinuities.\\
  
A key feature of the model is that all individual energy levels have
exact analytical expressions in
terms of $t$, $U$, and $\Delta v_{\rm ext}$
~\cite{carrascal2015hubbard,deur2017exact,senjean2018unified,Cernatic2022},
thus allowing for the evaluation (to any arbitrary numerical accuracy)
of ensemble density-functional Hxc energies. More precisely, the
e$N$-centered ensemble noninteracting kinetic energy and
the ensemble exact Hartree-exchange (simply denoted EEXX) functionals have the following
exact closed-form expressions:
\begin{subequations}
\begin{align}
\label{eq:eNc_Ts}
T_{\rm s}^{\bxi}(n) &=  -2t\sqrt{(1-\xi)^{2}-(1-n)^{2}}
\\
\label{eq:eNc_Ex}
E_{\rm Hx}^{\bxi}(n)
&=\frac{U}{2}
\left[  1 + \xi - \frac{\xi_-}{2} 
+ \left(1 - 3\xi -
\frac{\xi_-}{2}\right)\left(\dfrac{1-n}{1-\xi}\right)^2\right].
\end{align}
\end{subequations}
\rev{As readily seen from Eq.~(\ref{eq:eNc_Ts}), $T_{\rm s}^{\bxi}(n)$
does not vary with the weight $\xi_-$ assigned to the ground ionized
state, unlike the EEXX functional, which exhibits such a variation
through the weight $\xi_0(\bxi)$ assigned to the
two-electron ground state [see Eq.~(\ref{eq:dF_dU_zero_HD_appendix})].
This is a consequence of the fact that the
KS orbitals occupancies within the ensemble are independent of $\xi_-$ [see
Eqs.~(\ref{eq:occ_homo_HD}) and (\ref{eq:occ_lumo_HD}), and
Eq.~(\ref{eq:TS-LFTransform_Hdim}) from the
Appendix~\ref{sec:appendix_exact_functionals}].}
Note that EEXX will refer later on to the following EEXX-only approximation,
\be
E_{\rm Hxc}^{\bxi}(n)\overset{\rm EEXX}{\approx} E_{\rm Hx}^{\bxi}(n).
\ee
The missing ensemble density-functional correlation energy $E_{\rm c}^{\bxi}(n)=F^{\bxi}(n) -
T_{\rm s}^{\bxi}(n)-E_{\rm Hx}^{\bxi}(n)$ [see Eq.~(\ref{eq:ensemble_Hxc_formal_exp})] can be evaluated from the
e$N$-centered ensemble extension $F^{\bxi}(n)$ of Lieb's
functional~\cite{LFTransform-Lieb,deur2017exact,senjean2018unified}. The
detailed computation of the exact ensemble Hxc density functional and its 
ensemble weight derivatives is discussed in further detail in Appendix \ref{app:Hdim_details}.

\rev{On that basis, the exact ensemble Hxc potential can be
computed within the Hubbard dimer model, as further discussed in the
following. First, the Hxc potential
difference (between sites 1 and 0) is obtained from the exact ensemble density-functional KS
potential difference [see Eq.~(\ref{eq:eNc_KS_dens_fun_pot_HD})]. The
last step consists of shifting the values of the Hxc potential on
sites 0 and 1 (see Eq.~(\ref{eq:shifting_pot_HD})) such that the exact Koopmans' theorem of the key Eq.~(\ref{eq:exact_Koopmans_theo_eNc-eDFT}) is
fulfilled, either for the two-electron ground state ($\nu=0$ in this
case and the associated weight is $\xi_0(\bxi)$) or the first (singlet) neutral excited
state ($\nu=1$ in this case and the
associated weight is
$\xi$). The computation of the appropriate shift requires the evaluation
of the e$N$-centered ensemble density-functional Hxc energy [first term on
the right-hand side of Eq.~(\ref{eq:exact_Koopmans_theo_eNc-eDFT})], which has
been discussed previously, as well as the derivatives with respect to
$\xi_-$ and $\xi$ of the ensemble Hxc functional [second and third terms on 
the right-hand side of Eq.~(\ref{eq:exact_Koopmans_theo_eNc-eDFT}),
where $\xi_1^N=\xi$]. The correlation part of the latter derivatives can
be determined {\it via} the
Hellmann--Feynman theorem from the variational Legendre--Fenchel
expression of $F^{\bxi}(n)$ [see Eqs.~(\ref{eq:deriv_Hxc_func_xi_HD_HellFeyn}) and
(\ref{eq:deriv_Hxc_func_xi_minus_HD_HellFeyn}) in Appendix~\ref{sec:appendix_exact_functionals}].}

\subsection{Observation of derivative discontinuities}

Let us first summarize the key findings of Appendix
\ref{app:Hdim_details}. For a given choice of $U$, $t$, and $\Delta
v_{\rm ext}$, as well as given ensemble weight values $\bxi$, the
ensemble Hxc potential difference (between sites 1 and 0) can be
expressed exactly as follows,  
\be\label{eq:eNc_Hxc_pot_HD}
\Delta v_{\rm Hxc}^{\bxi}=\Delta v_{\rm KS}^{\bxi}(n^{\bxi})-\Delta
v_{\rm ext},
\ee
where the ensemble density-functional KS potential difference (see
Appendix \ref{app:Hdim_details} and Ref.~\onlinecite{deur2017exact}),
\be\label{eq:eNc_KS_dens_fun_pot_HD}
\Delta v_{\rm KS}^{\bxi}(n)=\dfrac{\partial T_{\rm
s}^{\bxi}(n)}{\partial n}=\dfrac{2t(n-1)}{\sqrt{(1-\xi)^{2} -
(1-n)^{2}}},
\ee
is a {\it continuous} function of both the density $n$ (in the range $\xi \leq n\leq 2-\xi$) and the ensemble
weight $\xi$ assigned to the neutral excited state, and $n^{\bxi}$
denotes the true interacting ensemble density (on site 0). The ensemble Hxc
potential on site $i$ can be expressed as follows [see
Eq.~(\ref{eq:shifting_pot_HD})],  
\be\label{eq:on_site_eNc_Hxc_pot}
v_{{\rm Hxc},i}^{\bxi}= (-1)^{i-1}\dfrac{\Delta v_{\rm
Hxc}^{\bxi}}{2}-\mu^{\bxi}_{\rm Hxc},\quad i=0,1, 
\ee
where the uniform shift $\mu^{\bxi}_{\rm Hxc}$ is determined, for a given ionization
process, from Eq.~(\ref{eq:exact_Koopmans_theo_eNc-eDFT}).\\

We now turn to the results and their discussion. All the calculations, exact and approximate, have been carried out with $t=1/2$.
For convenience, EEXX results have been generated without density-driven
errors. In other words, exact interacting ensemble densities have been
employed in conjunction with Eq.~(\ref{eq:eNc_Hxc_pot_HD}). Therefore, the EEXX
approximation has been used only when evaluating $\mu^{\bxi}_{\rm Hxc}$
through the neglect of the ensemble correlation density functional and its weight
derivatives in Eq.~(\ref{eq:exact_Koopmans_theo_eNc-eDFT}).

\begin{figure}[htb]
\hspace*{-0.5cm}
\includegraphics[scale=0.5]{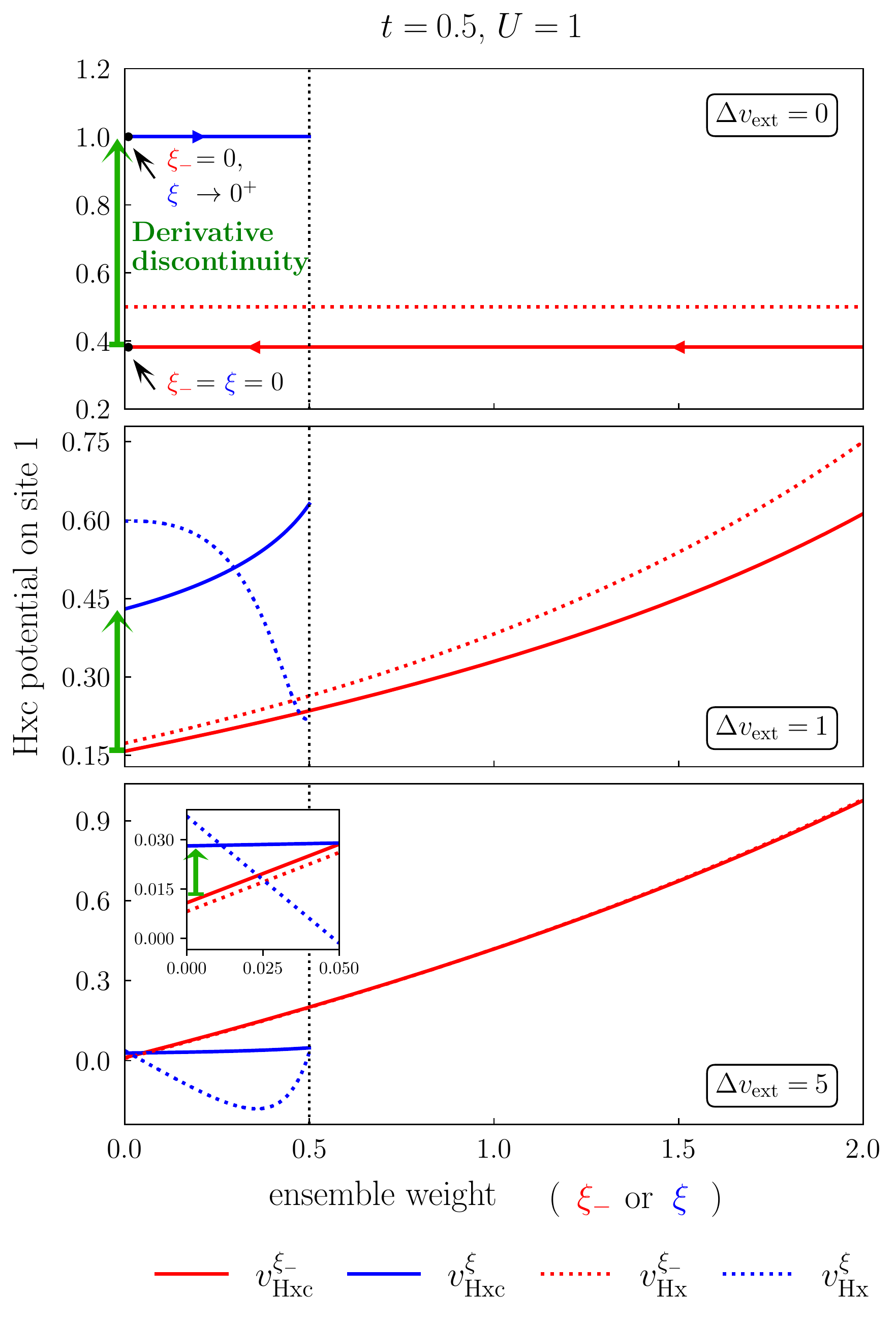}
\caption{Exact Hxc and EEXX-only (Hx) potentials on site 1 of the
Hubbard dimer plotted as
functions of the ensemble weights [$\xi_-$ for the ionization (in red) of the
two-electron ground state and $\xi$ for the neutral excitation process
(in blue)] for $U=2t=1$ and various external potential values. The derivative discontinuity that occurs
when switching from one excitation process to the other is
highlighted in green. The vertical black line at $\xi=1/2$ (dotted) indicates the TGOK equiesemble.}
\label{fig:DDiscont_Hdim}
\end{figure}

Figure \ref{fig:DDiscont_Hdim} shows the variation of the ensemble
Hxc potential on site 1 with respect to the ensemble weights for two
separate physical processes (see Appendix~\ref{app:DD_Hdim_general} for
the detailed derivation of the potential in both cases). The first one
(red curve in Fig.~\ref{fig:DDiscont_Hdim}) corresponds to the ionization of the two-electron ground
state ($0 \leq \xi_- \leq 2$ and $\xi = 0$ in this case). In the second
process (blue curve in Fig.~\ref{fig:DDiscont_Hdim}), the neutral excitation is included into the ensemble ($0 < \xi
\leq 1/2$) where the ionized ground state contributes
infinitesimally. In other words, ensemble weight
derivatives are evaluated for $\xi_-=0$ in this case. As readily seen
from Fig.~\ref{fig:DDiscont_Hdim}, the exact Hxc potential, constructed
according to Eq.~(\ref{eq:exact_Koopmans_theo_eNc-eDFT}), does exhibit a
discontinuity when switching from one process to the other, whether the
dimer is symmetric or not. As expected
from Eqs.~(\ref{eq:excitation_energy_from_TGOK-DFT}) and (\ref{eq:HxcDD_from_NceDFT_final_result}), this Hxc derivative discontinuity, which reduces to a xc derivative discontinuity if the regular
weight-independent Hartree functional is employed, equals
\be
\begin{split}
\Delta_{\rm Hxc}^0&:=v_{\rm Hxc,1}^{(\xi\rightarrow 0^+,0^+)}-v_{\rm Hxc,1}^{(\xi=0,0^+)}
\\
&= 
v_{\rm Hxc,0}^{(\xi\rightarrow 0^+,0^+)}-v_{\rm Hxc,0}^{(\xi=0,0^+)}
\\
&=\left.\dfrac{\partial E_{\rm Hxc}^{(\xi,0)}(n_{\Psi_0^N})}{\partial \xi}\right|_{\xi=0},
\end{split}
\ee
and corresponds to the deviation of the regular KS gap from its exact
interacting optical counterpart.
\\

As readily seen from
Eqs.~(\ref{eq:eNc_Hxc_pot_HD})--(\ref{eq:on_site_eNc_Hxc_pot}), the derivative discontinuity is
solely induced by the readjustment of the constant shift in the Hxc
potential when moving from the ionization of the ground state to the
neutral excitation process, {\ie},
\be
\Delta_{\rm Hxc}^0
=
\mu_{\rm Hxc}^{(\xi=0,0^+)}-\mu_{\rm
Hxc}^{(\xi\rightarrow 0^+,0^+)}.
\ee
This appears clearly in 
the symmetric dimer, for which $n^{\bxi}=1$, thus leading to $\Delta v_{\rm Hxc}^{\bxi}=\Delta v_{\rm
ext}=0$ and, therefore, $v_{{\rm Hxc},i}^{\bxi}=-\mu^{\bxi}_{\rm Hxc}$. In this
special case, the e$N$-centered ensemble Hx and correlation energies
read [see Eqs.~(\ref{eq:eNc_Ex}) and (\ref{eqappendix:eNc_correlation_symmetric})]
\begin{subequations}
\begin{align}
\left.E^{\bxi}_{\rm
Hx}(n)\right|_{n=1}& =\dfrac{U}{2}\left(1+\xi-\dfrac{\xi_-}{2}\right),
\\
\label{eq:Nce_corr_ener_symmetric_HD}
\left.E^{\bxi}_{\rm
c}(n)\right|_{n=1}& =\left(1-\xi-\dfrac{\xi_-}{2}\right)E_{\rm c},
\end{align}
\end{subequations}
respectively, where $E_{\rm c}=2t-\frac{1}{2}\sqrt{U^2 + 16t^2}$ is the regular ground-state correlation energy
of the symmetric two-electron Hubbard dimer. Consequently, for the
ionization of the excited state, the Hx and correlation contributions to
the right-hand side of Eq.~(\ref{eq:HxcDD_from_NceDFT_final_result}) can
be simplified as follows (we recall that $N=2$), 
\begin{subequations}
\begin{align}
\label{eq:Hx_contribution_Koopmans_neutralX}
E^{\bxi}_{\rm
Hx}(n)-(N+\xi_-)\dfrac{\partial E^{\bxi}_{\rm
Hx}(n)}{\partial \xi_-}+(N-\xi)\dfrac{\partial E^{\bxi}_{\rm
Hx}(n)}{\partial \xi}
& \overset{n=1}{=}2U,
\\
\label{eq:neutal_exc_no_corr_pot}
E^{\bxi}_{\rm
c}(n)-(N+\xi_-)\dfrac{\partial E^{\bxi}_{\rm
c}(n)}{\partial \xi_-}+(N-\xi)\dfrac{\partial E^{\bxi}_{\rm
c}(n)}{\partial \xi} 
& \overset{n=1}{=}0,
\end{align}
\end{subequations}
respectively. If, instead, we consider the ionization of the 
two-electron
ground state, we obtain
\begin{subequations}
\begin{align}
\left[E^{\bxi}_{\rm
Hx}(n)-(N+\xi_-)\dfrac{\partial E^{\bxi}_{\rm
Hx}(n)}{\partial \xi_-}\right]_{\xi=0}&\overset{n=1}{=}U
\\
\left[E^{\bxi}_{\rm
c}(n)-(N+\xi_-)\dfrac{\partial E^{\bxi}_{\rm
c}(n)}{\partial \xi_-}\right]_{\xi=0}&\overset{n=1}{=}2E_{\rm c},
\end{align}
\end{subequations}respectively. As a result, we deduce the following weight-independent
Hxc potential expressions (with the same value on both sites) for each excitation process (see
Appendix~\ref{app:DD_symmetric_Hdim}), $v_{\rm
Hxc}^{(0,\xi_-)}=(U+4t-\sqrt{U^2 + 16t^2})/2$ with $\xi_->0$
and $v_{\rm Hxc}^{(\xi,0^+)}=U$ with $\xi>0$, respectively. Thus we
recover the derivative discontinuity expression 
of Ref.~\onlinecite{deur2017exact}, \ie, $\Delta_{\rm Hxc}^{0}=(U-4t+\sqrt{U^2 +
16t^2})/2$. As the ratio $\Delta v_{\rm ext}/U$ (and, therefore, the
asymmetry of the model) increases for a fixed $U$ value, the derivative discontinuity decreases (see the central and bottom
panels of Fig.\ref{fig:DDiscont_Hdim}), as expected for a system where
electron correlation becomes weaker.
\\

Turning finally to the EEXX-only approximation, we obtain for the symmetric dimer the exact ensemble Hxc potential associated to the neutral
excitation, $v_{\rm Hx}^{(\xi,0^+)}=v_{\rm Hxc}^{(\xi,0^+)}=U$, as a
consequence of Eq.~(\ref{eq:neutal_exc_no_corr_pot}).
However, EEXX overestimates the $N$-centered
ensemble 
Hxc potential associated to the ionization of the two-electron ground
state, giving $v_{\rm Hx}^{(0,\xi_-)}=U/2$,
so that the Hx contribution to the derivative discontinuity is $\Delta_{\rm Hx}^{0}=U/2$. These results are in perfect
agreement with the top panel of Fig.~\ref{fig:DDiscont_Hdim}.
In the asymmetric case, EEXX is still able to model (approximately) the
derivative discontinuity, because the EEXX functional is ensemble weight-dependent. When
comparing the EEXX-only and exact Hxc ensemble potentials that describe
the neutral excitation, in the particular case where $U=\Delta v_{\rm ext}=1$ (see the solid and dotted blue lines in the
central panel of Fig.~\ref{fig:DDiscont_Hdim}), it becomes clear that electron
correlation can drastically change the variation in the TGOK ensemble
weight $\xi$ of the Hxc potential. In this specific correlation regime, the Hx
potential decreases drastically as $\xi$ approaches $1/2$, unlike the
full Hxc potential. This feature originates from the Hx contribution to
Koopmans' theorem [see Eq.~(\ref{eq:exact_Koopmans_theo_eNc-eDFT})],
from which the Hx potential is uniquely defined. This contribution
reads as in the left-hand side of
Eq.~(\ref{eq:Hx_contribution_Koopmans_neutralX}) and can be simplified as
follows, according to Eq.~(\ref{eq:eNc_Ex}),      
\be\label{eq:EEXX_cont_Koopmans_simpler}
\begin{split}
&E^{\bxi}_{\rm
Hx}(n)-(N+\xi_-)\dfrac{\partial E^{\bxi}_{\rm
Hx}(n)}{\partial \xi_-}+(N-\xi)\dfrac{\partial E^{\bxi}_{\rm
Hx}(n)}{\partial \xi}
\\
&\overset{\bxi=(\xi,0)}{=}U\left[2+\dfrac{\xi(1-n)^2(3\xi-5)}{(1-\xi)^3}\right]
=:\mathcal{D}^\xi_{\rm Hx}(n)
.
\end{split}
\ee
By expressing the Hx potential (on site 1) as follows, according to
Eqs.~(\ref{eq:eNc_Hxc_pot_HD}) and (\ref{eq:on_site_eNc_Hxc_pot}),
\be
v^{\xi}_{{\rm Hx}}:=v^{(\xi,0^+)}_{{\rm Hx},1}=\dfrac{1}{2}\left(\Delta v_{\rm KS}^{\xi}(n^{\xi})-\Delta
v_{\rm ext}\right)-\mu^{\xi}_{{\rm Hx}},
\ee
where, according to Koopmans' theorem of
Eq.~(\ref{eq:exact_Koopmans_theo_eNc-eDFT}) and the notation introduced
in Eq.~(\ref{eq:EEXX_cont_Koopmans_simpler}),  
\be
\left(\Delta v_{\rm KS}^{\xi}(n^{\xi})-\Delta
v_{\rm ext}\right)\left(1-n^{\xi}\right)
-2\mu^{\xi}_{{\rm Hx}}=\mathcal{D}^\xi_{\rm Hx}(n^{\xi}),
\ee
we obtain the final expression:
\be\label{eq:final_exp_Hx_pot_with_mu_Hx}
v^{\xi}_{{\rm Hx}}=\dfrac{n^{\xi}}{2}\left(\Delta v_{\rm KS}^{\xi}(n^{\xi})-\Delta
v_{\rm ext}\right)+\dfrac{\mathcal{D}^\xi_{\rm Hx}(n^{\xi})}{2}.
\ee

When $U=\Delta v_{\rm ext}=1$, the ensemble density varies weakly with
$\xi$ in the range $1.4\leq n^{\xi}\leq 1.5$~\cite{deur2017exact}, so that the 
contribution $\mathcal{D}^\xi_{\rm Hx}(n^{\xi})$ to the Hx potential varies essentially as
$\xi(3\xi-5)/(1-\xi)^3$, which decreases with $\xi$ over the range
$0\leq\xi\leq 1/2$, in agreement with the
middle panel of Fig.~\ref{fig:DDiscont_Hdim}. On the other hand, in the
more pronounced $\Delta v_{\rm ext}/U=5$ asymmetric regime (see the
bottom panel of Fig.~\ref{fig:DDiscont_Hdim}), the ensemble
density varies as $n^{\xi}\approx 2-\xi$~\cite{deur2017exact}. Consequently, the prefactor
$(1-n^{\xi})^2\approx (1-\xi)^2$ in the second term of
$\mathcal{D}^\xi_{\rm Hx}(n^{\xi})$ [see
Eq.~(\ref{eq:EEXX_cont_Koopmans_simpler})] does now contribute to the
weight dependence of the Hx potential, thus leading to a variation as
$\xi(3\xi-5)/(1-\xi)$ and, therefore, a substantially
different behavior (still decreasing though) when approaching $\xi=1/2$. The increase of the Hx
potential observed in the vicinity of $\xi=1/2$ (see the bottom panel of
Fig.~\ref{fig:DDiscont_Hdim}) can be related to the KS and external
potentials as well as to the variation in $\xi$ of the ensemble density [see
Eqs.~(\ref{eq:eNc_KS_dens_fun_pot_HD}) and
(\ref{eq:final_exp_Hx_pot_with_mu_Hx})]. Indeed, in the limit {$\Delta v_{\rm
ext}/(2t)=\Delta v_{\rm
ext}/U\rightarrow +\infty$, which is taken in order to better understand the strongly
asymmetric $\Delta v_{\rm ext}/U=5$ regime depicted in the bottom panel of Fig.~\ref{fig:DDiscont_Hdim}}, the ensemble density is essentially a
noninteracting one,
\be
n^\xi\approx 1+\frac{(1-\xi)\Delta v_{\rm ext}}{\sqrt{(\Delta v_{\rm
ext})^2+4t^2}}\approx 2-\xi-\dfrac{2(1-\xi)t^2}{(\Delta v_{\rm
ext})^2},  
\ee
which can, in this form, be introduced into the KS density-functional
expression of Eq.~(\ref{eq:eNc_KS_dens_fun_pot_HD}), thus leading to the
Hx potential contribution  
\be
\dfrac{n^{\xi}}{2}\left(\Delta v_{\rm KS}^{\xi}(n^{\xi})-\Delta
v_{\rm ext}\right)\approx -\dfrac{t^2}{\Delta
v_{\rm ext}}(2-\xi),
\ee
which increases with $\xi$. 

\subsection{Connection with regular ground-state DFT}

In order to establish a clearer connection with
conventional $N$-electron ground-state DFT ({\ie}, the $\bxi={\bf 0}$
limit of the theory), we adopt in this section a different perspective by using
the general and exact IP expressions of
Eq.~(\ref{eq:general_exp_I_nu_eNc_eDFT}), where the ensemble Hxc
potential is defined up to a constant. Unlike in
Sec.~\ref{sec:DDs_revisited}, and the previous section,
we do {\it not} impose any constraint on the latter constant. According
to Eq.~(\ref{eq:general_exp_I_nu_eNc_eDFT}) 
[see also Eqs.~(\ref{eq:IP_ground_Hdim}) and
(\ref{eq:IP_excited_Hdim})], the IPs of the
$N$-electron ($N=2$ here) ground and first neutral singlet excited
states can be expressed as follows when $\bxi={\bf 0}$, 
\begin{subequations}
\begin{align}
\label{eq:IP0_zeroweight_limit}
I_{0}^{N} & = -\tilde{\varepsilon}_{\rm H} +
\left.\dfrac{\partial E_{\rm Hxc}^{\xi_-}[n_{\Psi_0^N}]}{\partial
\xi_-}\right|_{\xi_-=0},
\\
\label{eq:IP1_zeroweight_limit}
I_{1}^{N} & = -\tilde{\varepsilon}_{\rm L} + 
\left.\dfrac{\partial E_{\rm Hxc}^{\bxi}[n_{\Psi_0^N}]}{\partial
\xi_-}\right|_{\xi_-=0} - \left.\dfrac{\partial E_{\rm
Hxc}^{\bxi}[n_{\Psi_0^N}]}{\partial \xi}\right|_{\xi=0},
\end{align}
\end{subequations}
respectively, in terms of the regular (defined up to a constant) KS HOMO
and LUMO energies ${\varepsilon}_i$ to which we have applied the
Levy--Zahariev (LZ) density-functional shift in potential~\cite{levy2014ground} 
\be
C[n]=\dfrac{E_{\rm
Hxc}[n]-\int d\br\, v_{\rm
Hxc}(\br)n(\br)}{N},
\ee
where $E_{\rm Hxc}[n]$ and $v_{\rm Hxc}(\br)$ are the regular Hxc
functional and potential, respectively: 
\be
{\varepsilon}_i\rightarrow \tilde{\varepsilon}_i={\varepsilon}_i+C[n_{\Psi_0^N}].
\ee 
The nice feature of LZ-shifted
orbital energies is that they are invariant under constant shifts in the
Hxc potential. In the present two-electron Hubbard dimer
model [see Eqs.~(\ref{eq:eNc_KS_dens_fun_pot_HD}),
\eqref{eq:KS_Hamiltonian_Hdim}, and
\eqref{eq:eHOMO_Hdim}], the unshifted KS energies are $\varepsilon_{\rm
H}=-\varepsilon_{\rm L}\equiv \varepsilon_{\rm
H}(\Delta v_{\rm KS}^{\bxi={\bf 0}}(n_{\Psi_0^N}))$  
and 
\be\label{eq:Hdim_LZShift}
C(n) = \dfrac{1}{2}[E_{\rm Hxc}(n) - (1-n)\Delta v_{\rm
Hxc}(n)].
\ee

\begin{figure}[htb]
\hspace*{-0.5cm}
\includegraphics[scale=0.5]{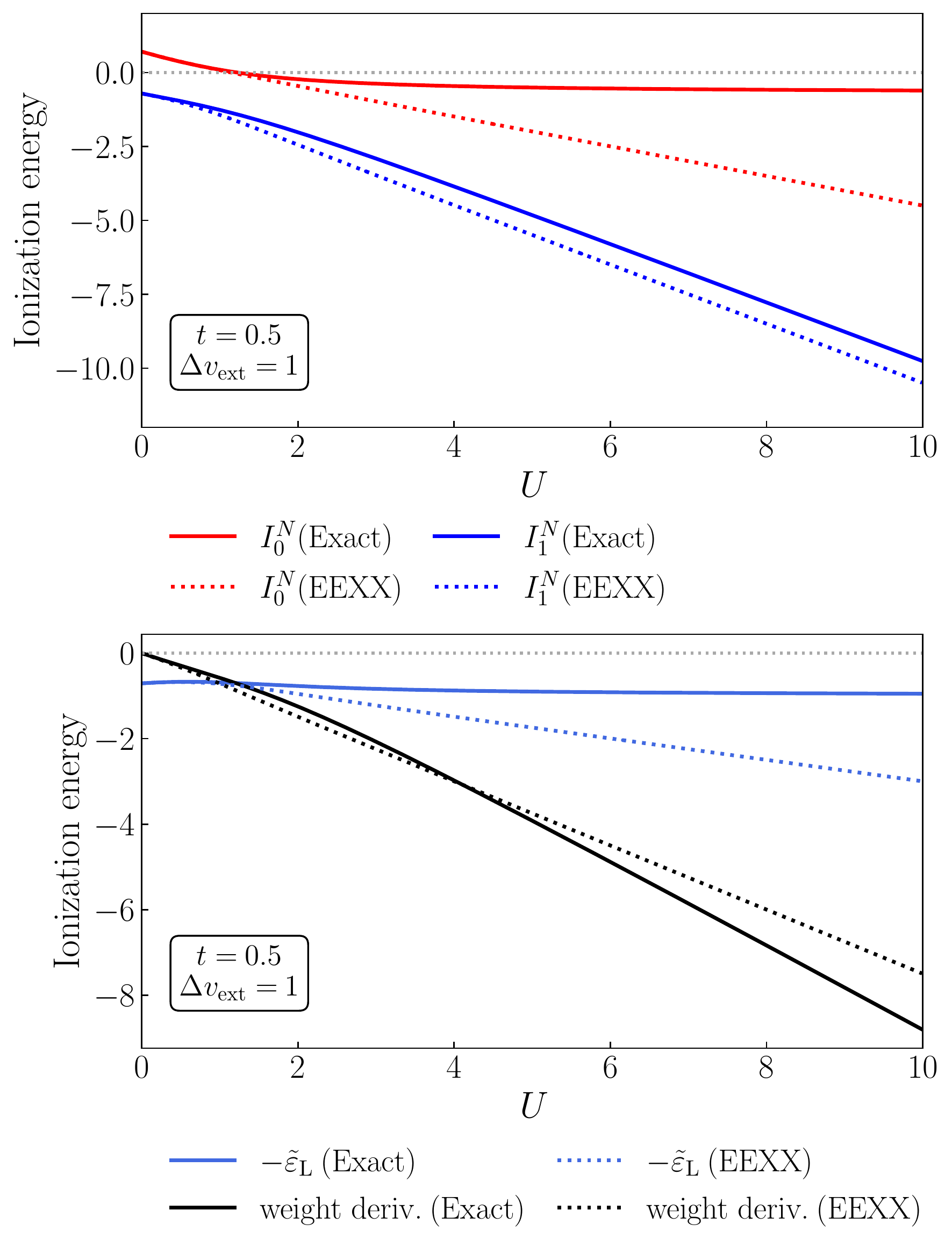}
\caption{Top: Variation of ground-state (red) and excited-state (blue) IPs with the
on-site repulsion $U$ in the
asymmetric two-electron Hubbard dimer ($N=2$ and $\Delta v_{\rm
ext}/t=2$). EEXX-only results (dotted lines)
are shown for analysis purposes. 
Bottom: Decomposition of the excited-state IP into LZ-shifted KS LUMO
energy and ensemble weight derivatives contributions [see Eq.~(\ref{eq:IP1_zeroweight_limit})].}
\label{fig:IPN01_bigu}
\end{figure}

The top panel of Fig.~\ref{fig:IPN01_bigu} shows the variations in $U$
of the ground-state and 
first excited-state IPs at the exact and approximate EEXX levels of
calculation in the asymmetric $\Delta v_{\rm ext}/t=2$ case. 
The trends of the ground-state IP in Fig.~\ref{fig:IPN01_bigu} are
qualitatively the same as in Fig.~4 of
Ref.~\onlinecite{senjean2018unified} (where $\Delta v_{\rm ext}/t=5$).
As long as electron correlation is weak or moderately strong ($U/t\leq
2$), EEXX performs very well.  
In the stronger correlation regime, the relatively good performance of
EEXX in the description of the excited-state IP was expected since, as
readily seen from the following energy expressions~\cite{deur2017exact}, 
\begin{subequations}
\begin{align}
\dfrac{E_0^N}{U}&=
\dfrac{4}{\qty(\frac{\Delta v_{\rm ext}}{U})^2-1}\left(\dfrac{t}{U}\right)^2
+\mathcal{O}\left[\left(\dfrac{t}{U}\right)^3\right],
\\
\dfrac{E_1^N}{U}&=1-\dfrac{\Delta v_{\rm ext}}{U}
+\dfrac{2}{1-\frac{\Delta v_{\rm ext}}{U}}\left(\dfrac{t}{U}\right)^2
+\mathcal{O}\left[\left(\dfrac{t}{U}\right)^3\right],
\end{align}
\end{subequations}
for $U\gg \Delta v_{\rm ext}\gg t$, the first (singlet) excited state
has essentially no
correlation energy. This
is not the case for the ground state, which explains the poor
performance of EEXX in the description of the ground-state IP when
electron correlation is strong.    

For completeness (the following analysis has already been done for the
ground-state IP in Ref.~\onlinecite{senjean2018unified}), the
decomposition into LZ-shifted KS LUMO energy and
ensemble Hxc weight derivatives of the excited-state IP [see
Eq.~(\ref{eq:IP1_zeroweight_limit})] is plotted as a function of the
interaction strength $U$ in the bottom panel of
Fig.~\ref{fig:IPN01_bigu}. As $U$ increases, the Hxc
ensemble weight derivatives
dominate while the LZ-shifted LUMO energy essentially tends to the
hopping parameter $t$,
which is the expected result for a perfectly symmetric dimer. Indeed,
the ground-state density approaches the value $n=1$ when $U/\Delta v_{\rm
ext}$ becomes large~\cite{deur2017exact,deur2018exploring}.
Note that, at the EEXX level of approximation, the
LZ-shifted LUMO energy is overestimated while the (absolute value of the) total
weight derivatives contribution is underestimated, so that the errors mostly cancel out when
evaluating the total excited-state IP (see the top panel of
Fig.~\ref{fig:IPN01_bigu}). Let us stress that, in the present case, being able
to model the weight dependence of the ensemble density-functional
exchange energy is essential.     
\\

\begin{figure*}[htb]
\includegraphics[scale=0.6]{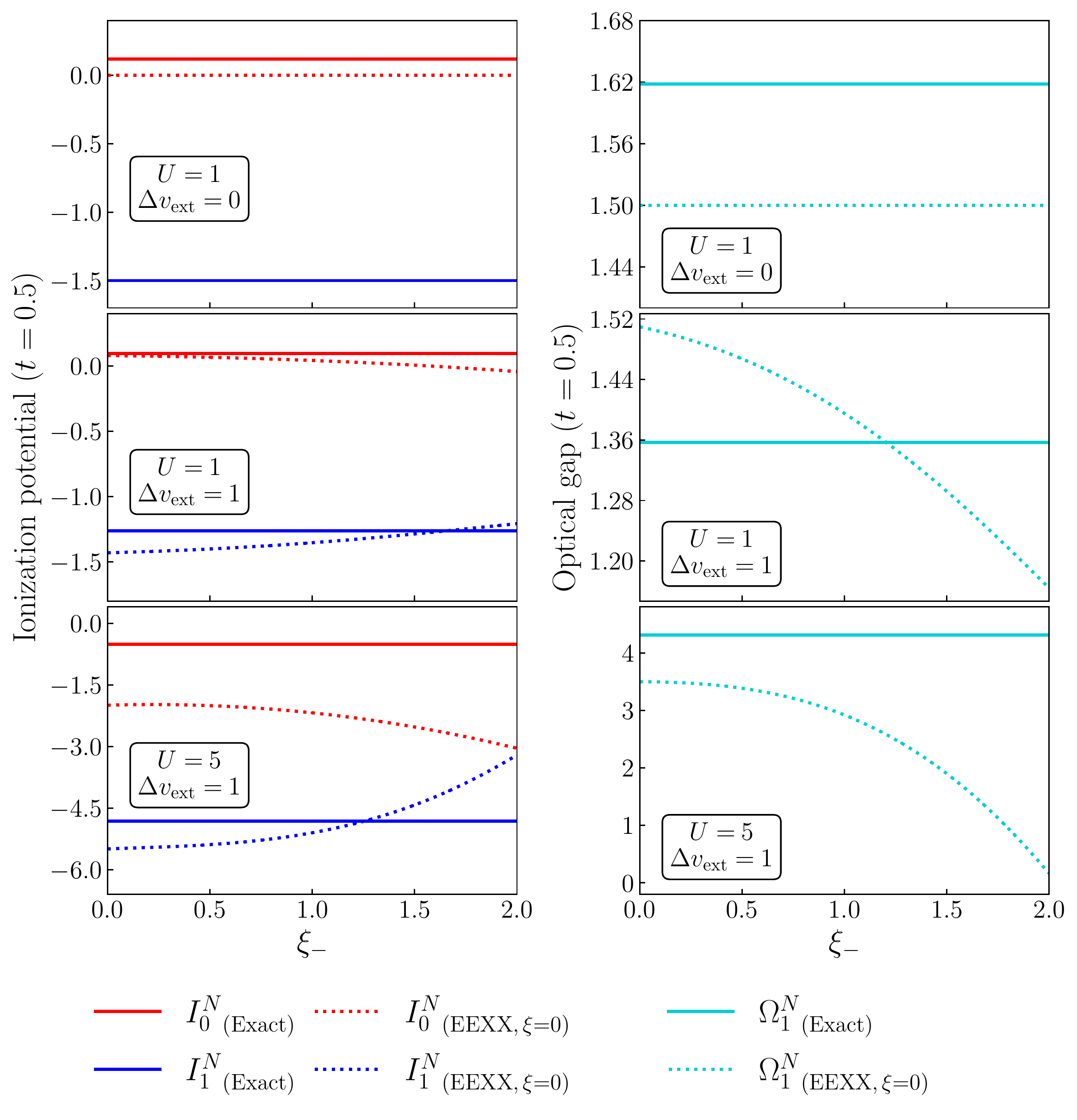}
\caption{Left panels: Exact and EEXX-only IPs for the ground state (red curves) and the excited state (blue curves) plotted as 
functions of the ionization ensemble weight $\xi_-$ in the limit
$\xi=0$ for the symmetric (top), and two asymmetric (middle, bottom)
Hubbard dimers. Right panels: Exact and EEXX-only optical gaps plotted
as the difference between the two IPs for the same three Hubbard dimers.}
\label{fig:IP_optgap_xim_w0}
\end{figure*}

For analysis purposes, we finally consider
deviations from regular ground-state DFT that are more than infinitesimal
by increasing the weight
assigned to the ionized ground state ($\xi_->0$) while remaining infinitesimally
close to the ground-state
limit (\ie, $\xi\rightarrow 0^+$) for the description of the excited
state. The impact on the evaluation of both ground- and excited-state
IPs, as well as the optical gap, is shown in
Fig.~\ref{fig:IP_optgap_xim_w0}. In the symmetric case (top panels),
$I_0^N=-t-\frac{1}{2}(U-\sqrt{U^2+16t^2})$, which gives the lower value
$I_0^N\approx t-\frac{U}{2}$ (\ie, $I_0^N\approx 0$ in
Fig.~\ref{fig:IP_optgap_xim_w0}, since $U=2t$)
within the EEXX-only approximation, where the correlation energy of the two-electron
ground state is missing. Since the exact excited-state IP $I^N_1=-t-U$ is recovered at the EEXX level in this case, the
underestimation of the optical gap is solely due to the
missing correlation in the ground state. 

Turning on asymmetry ($\Delta
v_{\rm ext}=1$) while remaining in the same moderately correlated regime
($U=2t$) introduces curvature in $\xi_-$ for both ground- and
excited-state approximate EEXX IPs (see the middle left panel of
Fig.~\ref{fig:IP_optgap_xim_w0}). Note that, in the present case ($U=\Delta
v_{\rm ext}=2t=1$), a perfect -- and probably fortuitous -- compensation of errors on
both IPs occurs when computing the EEXX optical gap for the specific 
weight value of $\xi_-\approx 1.3$. In the stronger correlation regime ($U=5$), the curvature
of EEXX IPs is further enhanced and perfect error cancellation does not occur
anymore (see the bottom panels of Fig.~\ref{fig:IP_optgap_xim_w0}).
Unlike in the exact theory, the EEXX density-functional corrections to
the bare {weight-dependent} KS IPs [see Eq.~(\ref{eq:general_exp_I_nu_eNc_eDFT})] do 
not lead to {weight-independent} physical IPs. Nevertheless, as shown analytically in Appendix~\ref{appendix:EEXX_largeU},
the EEXX functional succeeds in reversing, through its weight dependence, the variations in $\xi_-$ of both
ground- and first excited-state KS IPs, thus avoiding a further underestimation of the excited-state IP as
$\xi_-$ increases (see the bottom left panel of
Fig.~\ref{fig:IP_optgap_xim_w0}). As expected in such a strongly
correlated regime, a sensible optical gap can only be obtained by
retrieving the weight-dependent ensemble correlation
energy, which is completely neglected at the EEXX level of
approximation (see the bottom right panel of
Fig.~\ref{fig:IP_optgap_xim_w0}).

\section{Conclusions and outlook}\label{sec:conclu}

\rev{TGOK-DFT~\cite{JPC79_Theophilou_equi-ensembles,gross1988density} and the more recent $N$-centered
ensemble DFT~\cite{senjean2018unified,Senjean_2020} have been merged, thus allowing for an
in-principle exact and simultaneous density-functional description of both charged and
neutral electronic excitation processes [see the general
excitation energy expression of Eq.~(\ref{eq:any_XE_in_terms_KS_XE})]. The resulting unified theory is
referred to as extended $N$-centered ensemble DFT}. 
Unlike in conventional DFT, 
in extended $N$-centered ensemble DFT, the number of electrons is artificially held
constant and equal to the integer (so-called central) number of
electrons $N$ of the reference ground state, even when the ensemble under
study contains charged excited states in addition to neutral ones, hence the name of the theory.
Therefore, and most importantly, the Hxc potential is
always defined up to a constant shift within the present formalism.\\
  
The concept of derivative discontinuity for neutral excitations,
which is much less discussed in the literature~\cite{levy1995excitation,yang2014exact,gould2022single} than for charged
excitations, has been revisited in this context. When adjusting, for each
ground and excited $N$-electron states separately, the
constant shift in the Hxc potential, such that each of them satisfies an
exact ionization Koopmans' theorem [see
Eq.~(\ref{eq:exact_Koopmans_theo_eNc-eDFT})], a jump (the so-called derivative discontinuity)
is indeed observed when switching from the ground to the excited state
of interest [see Eq.~(\ref{eq:HxcDD_from_NceDFT_final_result})]. 

Unlike in previous works, the asymptotic behavior of the
density is never invoked nor used in the mathematical construction of
the derivative discontinuities. It fully relies, instead, on the weight dependence of the extended
$N$-centered ensemble Hxc density-functional energy. This allows for a
straightforward application of the theory to lattice models (the Hubbard
dimer in the present work). It may also open new perspectives in the
description of gaps in mesoscopic systems~\cite{Nazarov2021_Breakdown}. 
In principle, the present density-functional formalism is 
general enough to 
describe excitonic effects (through the inclusion of anionic states into
the ensemble), as well as the
challenging multiple-electron neutral excitation
processes. 

Finally, efforts should now be put into the design of weight-dependent density-functional
approximations, which is essential for turning the theory into a
reliable computational method. The (in-principle orbital-dependent)
ensemble exact-exchange functional that we studied within the Hubbard
dimer model nicely incorporates weight
dependencies but it obviously needs a proper (weight-dependent)
correlation counterpart whose construction is far from
trivial~\cite{loos2020weightdependent,PRL19_Gould_DD_correlation,Fromager_2020}. Exploring further the extension of
ensemble DFT to the
time-dependent linear response regime~\cite{PRL20_Gould_Hartree_def_from_ACDF_th}, for example, could be a
source of inspiration for this task. Work in these directions
is currently in progress.

\section*{Acknowledgements}
The authors thank ANR (CoLab project, grant no.: ANR-19-CE07-0024-02) for funding.
PFL thanks financial support from the European Research Council (ERC) under the European Union's Horizon 2020 research and innovation programme (Grant agreement No.~863481).

\appendix


\section{Technical details about the application of e$N$-centered ensemble DFT on the Hubbard dimer}\label{app:Hdim_details}

\subsection{Ensemble functionals}
In the following appendix, the exact and EEXX-approximate density functionals of the Hubbard dimer
are systematically derived for the e$N$-centered ensemble in Sec.~\ref{sec:Application_Hdim}.

\subsubsection{Exact functionals}\label{sec:appendix_exact_functionals}
We first consider the universal e$N$-centered ensemble density functional of the interacting system.
As stated in Sec.~\ref{subsec:Hdim_Key_features}, the central number of electrons was set to ${N=2}$.
For this particular choice, all individual $N$- and $(N-1)$-electron energies of the Hubbard dimer have analytical expressions,
which are available in, for example, Refs.~\onlinecite{deur2017exact,senjean2018unified}. 

The e$N$-centered ensemble energy reads as
\be\label{eq:eNc_interacting_energy_Hdim}
\begin{split}
E^{\bxi}(\Delta v_{\rm ext}) &= \left(1 - \dfrac{\xi_-}{2} - \xi\right)E_{0}^{N}(\Delta v_{\rm ext})
\\
&+\xi_-E_{0}^{N-1}(\Delta v_{\rm ext}) + \xi E_{1}^{N}(\Delta v_{\rm ext}),
\end{split}
\ee
where the dependence of individual energies on $t$ and $U$ has been omitted for clarity,
which will also be the case for other expressions dependent on $t$ and $U$.
The corresponding ensemble density reads as
\be\label{eq:eNc_interacting_density_Hdim}
\begin{split}
n^{\bxi}:=n^{\bxi}(\Delta v_{\rm ext}) &= \left(1 - \dfrac{\xi_-}{2} - \xi\right)n_{\Psi_0^N}(\Delta v_{\rm ext})
\\
&+\xi_-n_{\Psi_0^{N-1}}(\Delta v_{\rm ext}) 
+ \xi n_{\Psi_1^N}(\Delta v_{\rm ext}),
\end{split}
\ee
where individual $N$- and $(N-1)$-electron densities can be straightforwardly obtained from
first-order derivatives of individual energies with respect to the external potential. If we introduce
the counting operator $\hat{N} = \sum_{i=0}^{1}\hat{n}_{i}$,  it follows from Eq.~(\ref{eq:Hamiltonian_Hdim_vext})
and the Hellmann-Feynman theorem that
\be\label{eq:appendix_HD_density}
\begin{split}
\dfrac{\partial E_{I}^{M}(\Delta v_{\rm ext})}{\partial \Delta v_{\rm ext}}&=
\mybra{\Psi_{I}^{M}} 
\dfrac{\partial \hat{H}(\Delta v_{\rm ext})}{\partial \Delta v_{\rm ext}}
\myket{\Psi_{I}^{M}}
\\
&=
\mybra{\Psi_{I}^{M}} 
\dfrac{1}{2}(\hat{N} - 2\hat{n}_{0})
\myket{\Psi_{I}^{M}}
\\
&=\dfrac{M}{2}-n_{\Psi_{I}^{M}}(\Delta v_{\rm ext}),
\end{split}
\ee
where $M=N-1$ or $N$, and $I=0$ or $1$.

In the practical calculations reported in Sec.~\ref{sec:Application_Hdim}, we rely on the Legendre-Fenchel transform
definition of the universal e$N$-centered ensemble density functional, which can be obtained by
analogy with TGOK-DFT~\cite{deur2017exact} and $N$-centered eDFT~\cite{senjean2018unified},
\be\label{eq:F_LFTransform_Hdim}
F^{\bxi}(n) = \sup_{\Delta v}\biggl\{ E^{\bxi}(\Delta v) + \Delta v (n-1) \biggr\}.
\ee

In the above equation, the maximizing potential $\Delta v^{\bxi}(n)$ is equal to
the derivative of $F^{\bxi}(n)$ with respect to the density,
\be
\dfrac{\partial F^{\bxi}(n)}{\partial n} = \Delta v^{\bxi}(n).
\ee
Despite lacking a closed-form expression as a function of the density,
the value of $F^{\bxi}(n)$ can still be computed to arbitrary accuracy
for any given density from Eq.~(\ref{eq:F_LFTransform_Hdim}) via Lieb maximization. However, for the exact ensemble
density $n^{\bxi}$ which is used in all calculations in Sec.~\ref{sec:Application_Hdim}, 
the maximizing potential is simply equal to the external potential,
$\Delta v^{\bxi}(n^{\bxi}) = \Delta v_{\rm ext}$, and
$F^{\bxi}(n^{\bxi})=E^{\bxi}(\Delta v_{\rm ext})+\Delta v_{\rm ext}[n^{\bxi}(\Delta v_{\rm ext})-1]$
is an analytical function of $\Delta v_{\rm ext}$.

Next, we consider the noninteracting (KS) system which reproduces 
the e$N$-centered ensemble density of the
interacting system. From the KS Hamiltonian
\be\label{eq:KS_Hamiltonian_Hdim}
\hat{H}^{\rm KS}(\Delta v)=\hat{\mathcal{T}}+\frac{\Delta v}{2}(\hat{n}_1-\hat{n}_0),
\ee
we obtain the e$N$-centered ensemble energy as
\be
\begin{split}
\mathcal{E}_{\rm KS}^{\bxi}(\Delta v) &= \left(1-\xi-\frac{\xi_-}{2}\right)\mathcal{E}_{0}^{N}(\Delta v)
\\
&+{\xi_-}\mathcal{E}_{0}^{N-1}(\Delta v) +  \xi\mathcal{E}_{1}^{N}(\Delta v),
\end{split}
\ee
with the $N$- and ($N-1$)-electron ground-state energies, and the $N$-electron excited state energy written as
$\mathcal{E}_{0}^{N}(\Delta v) = 2\varepsilon_{\rm H}(\Delta v)$,  $\mathcal{E}_{0}^{N-1}(\Delta v) = \varepsilon_{\rm H}(\Delta v)$,
and $\mathcal{E}_{1}^{N}(\Delta v) = 0$, respectively, where
\be\label{eq:eHOMO_Hdim}
\varepsilon_{\rm H}(\Delta v) = -\sqrt{t^{2} + \dfrac{\Delta v^{2}}{4}}.
\ee

Similarly to the universal functional in Eq.~(\ref{eq:F_LFTransform_Hdim}), the
noninteracting kinetic energy $T^{\bxi}_{\rm s}(n)$ can be written as a Legendre-Fenchel transform, which reads
\be\label{eq:TS-LFTransform_Hdim}
T^{\bxi}_{\rm s}(n) = \sup_{\Delta v}\biggl\{ 2(1-\xi)\varepsilon_{\rm H}(\Delta v) - \Delta v (1-n) \biggr\}.
\ee
Since Eq.~(\ref{eq:TS-LFTransform_Hdim}) is explicitly independent of the ensemble
weight for the ($N-1$)-electron ground state $\xi_-$, it follows that the maximizing KS potential corresponding to the noninteracting e$N$-centered
ensemble is the same one as in TGOK-DFT~\cite{deur2017exact}. Hence, it can be expressed as follows,
\be\label{eq:deltavKS_Hdim}
\Delta v^{{\rm KS},\bxi}(n) \equiv  \Delta v^{{\rm KS},\xi}(n) 
= \dfrac{2t(n-1)}{\sqrt{(1-\xi)^{2} - (1-n)^{2}}},
\ee
which implies that a given density $n$ is noninteracting e$N$-centered
ensemble $v$-representable iff $\abs{n - 1} \leq 1 - \xi$. By plugging
Eq.~\eqref{eq:deltavKS_Hdim} into Eq.~\eqref{eq:TS-LFTransform_Hdim},
we obtain the expression for the noninteracting kinetic energy functional, $T_{\rm s}^{\bxi}(n)$, as
given in Eq.~\eqref{eq:eNc_Ts}. 

Finally, the ensemble weight derivatives of Hxc functional
$E_{\rm Hxc}^{\bxi}(n)=F^{\bxi}(n)-T_{\rm s}^{\bxi}(n)$, which are key ingredients
for obtaining exact expressions for the IPs and the optical gap, are obtained as follows,
\be\label{eq:deriv_Hxc_func_xi_HD_HellFeyn}
\begin{split}
\dxi{E_{\rm Hxc}^{\bxi}(n)} &= E_1^N(\Delta v^{\bxi}(n)) - E_0^N(\Delta v^{\bxi}(n))
\\
&-\dfrac{2t(1-\xi)}{\sqrt{(1-\xi)^2 - (1-n)^2}},
\end{split}
\ee
and
\be
\label{eq:deriv_Hxc_func_xi_minus_HD_HellFeyn}
\dxim{E_{\rm Hxc}^{\bxi}(n)} = E_0^{N-1}(\Delta v^{\bxi}(n)) 
- \dfrac{E_0^N(\Delta v^{\bxi}(n))}{2}.
\ee

\subsubsection{EEXX functional}
The e$N$-centered ensemble exact (Hartree) exchange (EEXX) functional
can be obtained from the formal expression of the Hx energy:
\be\label{eq:EHx_def_Hdim}
E_{\rm Hx}^{\bxi}(n)=
U\left.\dfrac{\partial F^{\bxi}(n)}{\partial U}\right|_{U=0},
\ee
where we have, according to Eq.~\eqref{eq:F_LFTransform_Hdim},
\be\label{eq:dF_dU_zero_HD_appendix}
\begin{split}
\left.\dfrac{\partial F^{\bxi}(n)}{\partial U}\right|_{U=0} &=
\left( 1 - \dfrac{\xi_-}{2} - \xi  \right)
\left.\dfrac{\partial E_0^N(\Delta v)}{\partial U}\right|_{U=0,\Delta v = \Delta v_{\rm KS}^{\bxi}(n)}
\\
&+\left.\xi\dfrac{\partial E_1^N(\Delta v)}{\partial U}\right|_{U=0,\Delta v = \Delta v_{\rm KS}^{\bxi}(n)},
\end{split}
\ee
and, according to Eqs.~\eqref{eq:eHOMO_Hdim} and \eqref{eq:deltavKS_Hdim}, and Eq.~(A7) in \cite{deur2017exact},
\begin{multline}\label{eq:EI_bigu_derivatives_Hdim}
\left.\dfrac{\partial E_I^N(\Delta v)}{\partial U}\right|_{U=0,\Delta v = \Delta v_{\rm KS}^{\bxi}(n)}
\\
= \dfrac{I+1}{2}\left[ 1 + \left(1-2I \right) \left(\dfrac{1-n}{1-\xi}\right)^2  \right],
\end{multline}
for $I=0$ or $1$. Finally, inserting Eq.~\eqref{eq:EI_bigu_derivatives_Hdim}
into Eq.~\eqref{eq:EHx_def_Hdim} gives, after some rearrangements of
various terms, the following expression for Hx energy of Eq.~(\ref{eq:eNc_Ex}) 
\be
E_{\rm Hx}^{\bxi}(n) = \dfrac{U}{2}\left[  1 + \xi - \dfrac{\xi_-}{2}
+ \left(1 - 3\xi -
\dfrac{\xi_-}{2}\right)\left(\dfrac{1-n}{1-\xi}\right)^2\right].
\ee

\subsection{Derivative discontinuity in the 
Hubbard dimer from the e$N$-centered ensemble perspective}

\subsubsection{General expressions}\label{app:DD_Hdim_general}

Within the formalism of extended $N$-centered ensembles,
the derivative discontinuity can be shown to appear even in the Hubbard dimer. Starting from
Eq.~\eqref{eq:general_exp_I_nu_eNc_eDFT}, we derive expressions for
the IPs of the ground and the first excited state,
\be\label{eq:IP_ground_Hdim}
\begin{split}
E^{N-1}_0 - E^{N}_0=
&-\varepsilon^{\bxi}_{\rm H}
\\
&-\dfrac{1}{2}\left(E_{\rm Hxc}^{\bxi}(n^{\bxi})-\sum_{i=0}^{1}v_{{\rm Hxc},i}^{\bxi} n^{\bxi}_i \right)
\\
&+\left(1+\dfrac{\xi_-}{2}\right)\left.\dxim{E_{\rm Hxc}^{\bxi}(n)}\right|_{n=n^{\bxi}}
\\
&+\dfrac{\xi}{2}\left.\dxi{E_{\rm Hxc}^{\bxi}(n)}\right|_{n=n^{\bxi}},
\end{split}
\ee
and
\be\label{eq:IP_excited_Hdim}
\begin{split}
E^{N-1}_0 - E^{N}_1=
&-\varepsilon^{\bxi}_{\rm L}
\\
&-\dfrac{1}{2}\left(E_{\rm Hxc}^{\bxi}(n^{\bxi})-\sum_{i=0}^{1}v_{{\rm Hxc},i}^{\bxi} n^{\bxi}_i \right)
\\
&+\left(1+\dfrac{\xi_-}{2}\right)\left.\dxim{E_{\rm Hxc}^{\bxi}(n)}\right|_{n=n^{\bxi}}
\\
&+\left(\dfrac{\xi}{2}-1\right)\left.\dxi{E_{\rm Hxc}^{\bxi}(n)}\right|_{n=n^{\bxi}},
\end{split}
\ee
where the integral $\int d\br v_{\rm Hxc}^{\bxi}(\br) n^{\bxi}(\br)$ has been replaced
by a summation over lattice sites.
For the purpose of displaying the derivative discontinuity, the KS
Hamiltonian of the Hubbard dimer is written with the KS potential
determined up to a constant as follows,
\be
\hat{H}^{{\rm KS},\bxi}=\hat{\mathcal{T}}+v^{{\rm KS},\bxi}_1\hat{n}_1+v^{{\rm
KS},\bxi}_0\hat{n}_0,
\ee
where
\begin{subequations}
\begin{align}
v^{{\rm KS},\bxi}_i & =\frac{(-1)^{i-1}}{2}\Delta v^{{\rm KS},\bxi}-\mu^{\bxi}_{\rm
Hxc},
\\
\Delta v^{{\rm KS},\bxi}&=\left.\partial T^{\bxi}_{\rm s}(n)/\partial
n\right|_{n=n^{\bxi}}
\equiv \Delta v_{\rm ext}+\Delta v^{\bxi}_{\rm Hxc},
\end{align}
\end{subequations}
and $n^{\bxi}$ is the exact e$N$-centered ensemble
occupation of site 0. The total Hxc potential on site $i$ reads
\be\label{eq:vHxc_i_Hdim}
v^{\bxi}_{{\rm Hxc},i}=\frac{(-1)^{i-1}}{2}\Delta v^{\bxi}_{\rm Hxc}-\mu^{\bxi}_{\rm
Hxc}.
\ee
The constant shift, $-\mu^{\bxi}_{\rm Hxc}$, is determined
to fulfill the constraint for exactification of Koopmans' theorem
in Eq.~\eqref{eq:exact_Koopmans_theo_eNc-eDFT}.

In order to derive the derivative discontinuity, we consider two specific ensembles.
The first ensemble is the regular $N$-centered ensemble, where the ensemble weights take the values of $\xi=0$ and $0\leq\xi_-\leq2$,
describing the ionization process from the two-electron ground state, with the
IP ${I_0^N = E_0^{N-1}-E_0^N}$.
To fulfill Koopman's theorem for the ground-state IP, we set all the terms on the right-hand sides
of Eq.~\eqref{eq:IP_ground_Hdim} to zero, except the HOMO energy.
This gives the following constraint for the Hxc potential in the Hubbard dimer,
\be
\sum_{i=0}^{1} v_{{\rm Hxc},i}^{(0,\xi_-)} n^{\xi_-}_i =
E^{\xi_-}_{\rm Hxc}(n^{\xi_-})-
\left(2+\xi_- \right)\left.\dxim{E_{\rm Hxc}^{\xi_-}(n)}\right|_{n=n^{\xi_-}},
\ee
where $n^{\xi_-}=n^{(0,\xi_-)}$ and $E_{\rm Hxc}^{\xi_-}=E_{\rm Hxc}^{(0,\xi_-)}$.
By inserting Eq.~\eqref{eq:vHxc_i_Hdim} into the above equation, and solving for $-\mu^{(0,\xi_-)}_{\rm Hxc}$, we get
\be
\begin{split}\label{eq:muHxc_ximinus_Hdim}
    -\mu_{\rm Hxc}^{(0,\xi_-)}&=\frac{1}{2}\left( \Delta v_{\rm Hxc}^{\xi_-}[n^{\xi_-}-1] + E_{\rm Hxc}^{\xi_-}(n^{\xi_- })\right)
\\
  &- \dfrac{2+\xi_-}{2}\left.\dfrac{\partial E_{\rm xc}^{\xi_-}(n)}{\partial \xi_-}\right|_{n = n^{\xi_-}},
\end{split}
\ee
where, according to Eqs.~\eqref{eq:eNc_Hxc_pot_HD} and \eqref{eq:eNc_KS_dens_fun_pot_HD},
$\Delta v_{\rm Hxc}^{\xi_-}=\Delta v_{\rm Hxc}^{(0,\xi_-)}(n^{\xi_-})=\Delta v_{\rm Hxc}^{(0,0)}(n^{\xi_-})$.

The second ensemble is the extended $N$-centered ensemble from Sec.~\ref{sec:Application_Hdim}, with the full weight range
for the neutral excitation ($0<\xi\leq1/2$), and the infinitesimal limit for the ionization
from the two-electron ground state, \textit{i.e.} $\xi_-\rightarrow 0^+$. The latter weight
limit is imposed so that Koopmans' theorem for the excited-state IP $I_1^N=E_0^{N-1}-E_1^N$ can still be fulfilled.
By setting all the terms on the right-hand side of Eq.~\eqref{eq:IP_excited_Hdim}
to zero, except the LUMO energy, we get
\be
\begin{split}\label{eq:exact_constraint_Hxc_pot_w_Hdim}
\sum_{i=0}^{1} v_{{\rm Hxc},i}^{(\xi,0^+)} n^{\xi}_i
&= E_{\rm Hxc}^{\xi}(n^{\xi}) + (2-\xi)\left.\dfrac{\partial E_{\rm xc}^{\xi}(n)}{\partial \xi}\right|_{n = n^{\xi}}
\\
&-2\left.\dfrac{\partial E_{\rm xc}^{(\xi,\xi_-)}(n^{\xi})}{\partial \xi_-}\right|_{\xi_- = 0},
\end{split}
\ee
where $n^{\xi}=n^{(\xi,0)}$ and $E_{\rm Hxc}^{\xi}=E_{\rm Hxc}^{(\xi,0)}$.
After inserting Eq.~\eqref{eq:vHxc_i_Hdim} into the above equation, and solving for $-\mu^{(\xi,0^+)}_{\rm Hxc}$, we get
\be
\begin{split}\label{eq:muHxc_xi_Hdim}
-\mu_{\rm Hxc}^{(\xi,0^+)}&=\frac{1}{2}\left( \Delta v_{\rm Hxc}^{\xi}[n^{\xi}-1] + E_{\rm Hxc}^{\xi}(n^{\xi})\right)
\\
&+
\dfrac{2-\xi}{2}\left.\dfrac{\partial E_{\rm xc}^{\xi}(n)}{\partial \xi}\right|_{n = n^{\xi}}
- \left.\dfrac{\partial E_{\rm xc}^{(\xi,\xi_-)}(n)}{\partial \xi_-}\right|_{\xi_-=0},
\end{split}
\ee
where $\Delta v_{\rm Hxc}^{\xi}=\Delta v_{\rm Hxc}^{(\xi,0)}(n^{\xi})$.
Considering now Eq.~\eqref{eq:muHxc_ximinus_Hdim}, the limit $\xi_-\rightarrow0^+$
determines the value of the constant Hxc shift for the two-electron ground state,
\be\label{eq:muHxc_ximinus_Hdim_GS}
\begin{split}
-\mu_{\rm Hxc}^{(\xi=0,0^+)} &= \dfrac{1}{2}\left(\Delta v_{\rm Hxc}[n_{\Psi_0^N}-1] +
E_{\rm Hxc}(n_{\Psi_0^N})\right)
\\
&- \left.\dfrac{\partial E_{\rm xc}^{\xi_-}(n_{\Psi_0^N})}{\partial \xi_-}\right|_{\xi_-=0}.
\end{split}
\ee
In the case of Eq.~\eqref{eq:muHxc_xi_Hdim},
the limit $\xi\rightarrow 0^{+}$ returns the value
of the Hxc shift for the TGOK ensemble with an infinitesimal amount of neutral excitation
from the two-electron ground state.
\be\label{eq:muHxc_xi_Hdim_GS}
\begin{split}
-\mu_{\rm Hxc}^{(\xi\rightarrow0^+,0^+)} &= \dfrac{1}{2}\left(\Delta v_{\rm Hxc}[n_{\Psi_0^N}-1] +
E_{\rm Hxc}(n_{\Psi_0^N})\right)
\\
&+ \left.\dfrac{\partial E_{\rm xc}^{\xi}(n_{\Psi_0^N})}{\partial \xi}\right|_{\xi=0}
- \left.\dfrac{\partial E_{\rm xc}^{\xi_-}(n_{\Psi_0^N})}{\partial \xi_-}\right|_{\xi_-=0}
\end{split}
\ee

By subtracting Eq.~\eqref{eq:muHxc_ximinus_Hdim_GS} from Eq.~\eqref{eq:muHxc_xi_Hdim_GS} and
using Eq.~\eqref{eq:vHxc_i_Hdim}, we obtain
the following expression for the derivative discontinuity,
\be
\begin{split}
v_{{\rm Hxc},1}^{(\xi\rightarrow 0^{+},0^+)} - v_{{\rm Hxc},1}^{(\xi=0,0^+)}
&=\mu_{\rm Hxc}^{(\xi=0,0^+)} - \mu_{\rm Hxc}^{(\xi\rightarrow 0^{+},0^{+})}
\\
&=\left.\dfrac{\partial E_{\rm xc}^{\xi}(n_{\Psi_0^N})}{\partial \xi}\right|_{\xi=0}.
\end{split}
\ee
This is an exact result for the derivative discontinuity in the Hubbard dimer, which is shown in Fig.~\ref{fig:DDiscont_Hdim}.

\subsubsection{Symmetric Hubbard dimer}\label{app:DD_symmetric_Hdim}

In the symmetric dimer, the Hxc potential and the derivative discontinuity have closed-form expressions.
Starting from the formulae for individual $N$- and $(N-1)$-electron energies, ${E_0^N(\Delta v_{\rm ext}=0)=\tfrac{1}{2}(U -
\sqrt{U^2 + 16t^2})}$, ${E_1^N(\Delta v_{\rm ext}=0)=U}$ and ${E_0^{N-1}(\Delta
v_{\rm ext}=0)=-t}$, the universal and the noninteracting kinetic
energy e$N$-centered ensemble density functionals are derived as follows,
\be\label{eq:F_sym_Hdim}
\begin{split}
F^{\bxi}(n=1) &= E^{\bxi}(\Delta v_{\rm ext}=0)
\\
&=\left( 1 - \dfrac{\xi_-}{2} - \xi \right)\dfrac{U - \sqrt{U^2 + 16t^2}}{2}
\\
&-\xi_- t + \xi U,
\\
\end{split}
\ee
and
\be
\label{eq:Ts_sym_Hdim}
T_{\rm s}^{\bxi}(n=1) = -2t(1-\xi),
\ee
while the EEXX functional reads
\be\label{eq:E_Hx_sym_Hdim}
E_{\rm Hx}^{\bxi}(n=1) = \dfrac{U}{2}\left(  1 - \frac{\xi_-}{2} + \xi \right).
\ee

For the regular $N$-centered ensemble ($\xi=0$ and ${0\leq\xi_-\leq2}$), the Hxc functional reads
\be
E_{\rm Hxc}^{\xi_-}(n=1) = \left( 1 - \dfrac{\xi_-}{2}\right)\dfrac{U - \sqrt{U^2 + 16t^2}}{2}
- \xi_- t + 2t,
\ee
which can also be expressed as
\be
E_{\rm Hxc}^{\xi_-}(n=1) = -(2 - \xi_-)\dfrac{\partial E_{\rm Hxc}^{\xi_-}(n=1)}{\partial \xi_-}.
\ee
Inserting the above expression into Eq.~\eqref{eq:muHxc_ximinus_Hdim},  and taking into account that $\Delta v_{\rm Hxc}^{\xi_-}=0$,
we obtain from Eq.~\eqref{eq:vHxc_i_Hdim} that
\be\label{eq:vHxc_ximinus_Hdim_symmetric}
\begin{split}
v_{{\rm Hxc},1}^{(0,\xi_-)} &= -2\dfrac{\partial E_{\rm Hxc}^{\xi_-}(n=1)}{\partial \xi_-}
\\
&=\frac{U+4t-\sqrt{U^2 + 16t^2}}{2}.
\end{split}
\ee
Using the EEXX functional [see Eq.~\eqref{eq:E_Hx_sym_Hdim})] instead of the exact Hxc functional
clearly gives the wrong result,
\be\label{eq:vHx_ximinus_Hdim_symmetric}
v_{{\rm Hx},1}^{(0,\xi_-)}=U/2.
\ee

Upon inclusion of the neutral excitation into the ensemble ($0<\xi\leq1/2$),
we observe that the EEXX-only and the exact Hxc ensemble potentials coincide
in the symmetric dimer. To demonstrate this, we first isolate the e$N$-centered ensemble correlation functional from
Eqs.~\eqref{eq:F_sym_Hdim},~\eqref{eq:Ts_sym_Hdim} and~\eqref{eq:E_Hx_sym_Hdim}:
\be\label{eqappendix:eNc_correlation_symmetric}
\begin{split}
E_{\rm c}^{\bxi}(n=1)&=F^{\bxi}(n=1) - T_{\rm s}^{\bxi}(n=1) - E_{\rm Hx}^{\bxi}(n=1)
\\
&=\left(  1 - \dfrac{\xi_-}{2} - \xi \right)\dfrac{4t - \sqrt{U^2 + 16t^2}}{2},
\end{split}
\ee
which can also be expressed as
\be
\begin{split}           
E_{\rm c}^{\bxi}(n=1) =                                       
&-(2-\xi)\dfrac{\partial E_{\rm c}^{\bxi}(n=1)}{\partial \xi}
\\                                                                   
&+ (2 + \xi_-)\dfrac{\partial E_{\rm c}^{\bxi}(n=1)}{\partial \xi_-}.
\end{split}
\ee
In the particular case when $\xi_-\rightarrow0^+$,  we have
\be
\begin{split}
E_{\rm c}^{\xi}(n=1)= 
&-(2-\xi)\dfrac{\partial E_{\rm c}^{\xi}(n=1)}{\partial \xi} 
\\
&+\left.2\dfrac{\partial E_{\rm c}^{(\xi_-,\xi)}(n=1)}{\partial \xi_-}\right|_{\xi_-=0}.
\end{split}
\ee

By inserting the above expression into Eq.~\eqref{eq:muHxc_xi_Hdim}, we can see that the correlation
part in the $E_{\rm Hxc}^{\bxi}(n=1)$ functional is canceled out by its weight derivatives, which
implies that only exchange terms contribute to $v_{\rm Hxc}^{(\xi,0^+)}$, giving the following result:
\be\label{eq:vHxc_xi_Hdim_symmetric}
\begin{split}
v_{{\rm Hxc},1}^{(\xi,0^+)}&\equiv v_{{\rm Hx},1}^{(\xi,0^+)}
\\
&=\dfrac{1}{2}\left( \dfrac{U}{2} + 2\dfrac{\partial E_{\rm Hx}^{\xi}(n=1)}{\partial \xi} \right)
-\left.\dfrac{\partial E_{\rm Hx}^{\bxi}(n=1)}{\partial \xi-}\right|_{\xi_-=0} 
\\
& = U.
\end{split}
\ee
Hence, according to Eqs.~\eqref{eq:vHxc_ximinus_Hdim_symmetric} and~\eqref{eq:vHxc_xi_Hdim_symmetric}, the exact derivative discontinuity in the symmetric dimer reads
\be
v_{{\rm Hxc},1}^{(\xi\rightarrow 0^{+},0^+)} - v_{{\rm Hxc},1}^{(\xi=0,0^+)}= \dfrac{U-4t+\sqrt{U^2 + 16t^2}}{2},
\ee
while the EEXX-only approximation, obtained from Eqs.~\eqref{eq:vHx_ximinus_Hdim_symmetric}
and~\eqref{eq:vHxc_xi_Hdim_symmetric}, reads
\be
v_{{\rm Hx},1}^{(\xi\rightarrow 0^{+},0^+)} - v_{{\rm Hx},1}^{(\xi=0,0^+)}= \dfrac{U}{2},
\ee
which is always lower than the exact derivative discontinuity.

\section{EEXX approximation applied to the Hubbard dimer with $U\gg \Delta
v_{\rm ext}=2t=1$}\label{appendix:EEXX_largeU}

In the strongly correlated regime $U\gg \Delta
v_{\rm ext}=2t=1$, the ground-state density is very close to
1~\cite{deur2017exact}, while the one-electron ground-state density
$n_{\Psi^{N-1}_0}$ (we recall that the central number of electrons is
$N=2$ in the present model)
fulfills, according to Eqs.~(\ref{eq:appendix_HD_density}) and (\ref{eq:eHOMO_Hdim}),
\be
\begin{split}
\dfrac{\partial \varepsilon_{\rm H}(\Delta
v_{\rm ext})}{\partial \Delta
v_{\rm ext}}&=-\dfrac{\Delta
v_{\rm ext}}{4\sqrt{t^2+(\Delta
v^2_{\rm ext}/4)}}
\\
&=\dfrac{1}{2}\qty(1-2n_{\Psi^{N-1}_0}),
\end{split}
\ee  
or, equivalently,
$n_{\Psi^{N-1}_0}=\frac{1}{2}(1+\frac{1}{\sqrt{2}})$, thus leading to
the following expression for the ensemble density under study
\be
\begin{split}
n^{\xi_-}&=n^{(\xi=0,\xi_-)}
\\
&={\left(1-\frac{\xi_-}{2}\right)\times
1+\xi_-\times \frac{1}{2}\left(1+\frac{1}{\sqrt{2}}\right)}
\\
&=1+\dfrac{\xi_-}{2\sqrt{2}}.
\end{split}
\ee
As a result, the ensemble KS potential varies with $\xi_-$ as follows
[see Eq.~(\ref{eq:eNc_KS_dens_fun_pot_HD})],
\be
\Delta v^{\xi_-}_{\rm
KS}=\dfrac{2t\left(n^{\xi_-}-1\right)}{\sqrt{1-\left(n^{\xi_-}-1\right)^2}}
=\dfrac{\xi_-}{\sqrt{8-\xi_-^2}},
\ee 
and the KS HOMO/LUMO energies read
\be
-\varepsilon_{\rm H}^{\xi_-}=\varepsilon_{\rm
L}^{\xi_-}=\sqrt{t^2+\frac{\left(\Delta v^{\xi_-}_{\rm
KS}\right)^2
}{4}}
=\dfrac{\sqrt{2}}{\sqrt{8-\xi_-^2}}.
\ee
Therefore, the bare KS ground-state (excited-state) IP increases
(decreases) with the ensemble weight $\xi_-$, unlike the approximate
interacting EEXX one (see the bottom left panel of
Fig.~\ref{fig:IP_optgap_xim_w0}). 

In order to identify the origin of
the latter trend we need to consider the decomposition of both IPs as
given in Eq.~(\ref{eq:general_exp_I_nu_eNc_eDFT}). The Hxc potential
term, which contributes to both IPs, reads
\be
\begin{split}
\dfrac{\Delta v^{\xi_-}_{\rm
Hxc}}{2N}\left(2-2n^{\xi_-}\right)&=-\dfrac{\xi_-\left(\Delta v^{\xi_-}_{\rm
KS}-\Delta v_{\rm ext}\right)}{4\sqrt{2}}
\\
&=\dfrac{1}{4\sqrt{2}}\left(-\dfrac{\xi_-^2}{\sqrt{8-\xi_-^2}}+\xi_-\right)
,
\end{split}
\ee
since $N=2$. Note that, as we approach the $\xi_-=2$ limit, the above
contribution 
decreases with $\xi_-$. Therefore, the weight dependence of the approximate
EEXX excited-state IP originates from the remaining ensemble Hx
energy term and its weight derivatives. The contribution that is common to both
ground- and excited-state IPs reads
\be
I^{\xi_-}_{0,\rm Hx}=\left[-\dfrac{E^{\xi_-}_{\rm
Hx}(n)}{N}+\left(1+\dfrac{\xi_-}{N}\right)\dfrac{\partial E^{\xi_-}_{\rm 
Hx}(n)}{\partial \xi_-}\right]_{n=n^{\xi_-}},
\ee
where [see Eq.~(\ref{eq:eNc_Ex})]
\be
E^{\xi_-}_{\rm
Hx}(n)=\dfrac{U}{2}\left(1-\dfrac{\xi_-}{2}\right)\left[1+(n-1)^2\right],
\ee               
thus leading to
\be\label{eq:GS_IP_Hx_contribution}
I^{\xi_-}_{0,\rm
Hx}=-\dfrac{U}{2}\left[1+(n^{\xi_-}-1)^2\right]
=-\dfrac{U}{2}\left(1+\dfrac{\xi^2_-}{8}\right).
\ee
The above contribution is therefore responsible for the net decrease of
the EEXX ground-state IP as the ensemble weight $\xi_-$ increases.\\

Turning to the excited-state IP, we need to consider the following
additional contribution [see Eq.~(\ref{eq:general_exp_I_nu_eNc_eDFT})]:
\be
\begin{split}
-\Delta_{\rm Hx}^{\xi_-}&=
-\left.\dfrac{\partial E^{(\xi,\xi_-)}_{\rm
Hx}(n)}{\partial \xi}\right|_{\xi=0,n=n^{\xi_-}}
\\
&=-\dfrac{U}{2}\left[1-(n^{\xi_-}-1)^2\left(1+\xi_-\right)\right]
\\
&=-\dfrac{U}{2}\left[1-\dfrac{\xi_-^2\left(1+\xi_-\right)}{
8}\right],
\end{split}
\ee
which, when it is added to the remaining Hx contribution of
Eq.~(\ref{eq:GS_IP_Hx_contribution}), gives the total Hx term
\be
I^{\xi_-}_{0,\rm Hx}-\Delta_{\rm
Hx}^{\xi_-}=-\dfrac{U}{2}\left(2-\dfrac{\xi_-^3}{8}\right).
\ee
The above contribution is therefore responsible for the net increase
with $\xi_-$ of the EEXX
excited-state IP. 

In summary, at the EEXX level of approximation and in
the considered $U\gg \Delta
v_{\rm ext}=2t=1$ regime, the ground- and first excited-state IPs read 
\begin{subequations}
\begin{align}
I^N_0\approx \dfrac{1}{4\sqrt{2}}\left(\xi_-+\sqrt{8-\xi_-^2}\right)-\dfrac{U}{2}\left(1+\dfrac{\xi^2_-}{8}\right),
\\
I^N_1\approx \dfrac{1}{4\sqrt{2}}\left(\xi_--\dfrac{8+\xi_-^2}{\sqrt{8-\xi_-^2}}\right)-\dfrac{U}{2}\left(2-\dfrac{\xi_-^3}{8}\right),
\end{align}
\end{subequations}
respectively, thus leading to the following expression for the optical
gap
\be
\begin{split}
I_0^N-I^N_1&\approx
\dfrac{2\sqrt{2}}{\sqrt{8-\xi_-^2}}
+\dfrac{U}{2}\left[1-\dfrac{\xi_-^2(1+\xi_-)}{8}\right].
\end{split}
\ee
As a result, we obtain in the limiting $\xi_-=0$ and $\xi_-=2$ cases the following expressions,
\begin{subequations}
\begin{align}
I^N_0&\overset{\xi_-=0}{\approx}\dfrac{1}{2} - \dfrac{U}{2},
\\ 
I^N_1&\overset{\xi_-=0}{\approx}-\dfrac{1}{2} + U,
\\
I^N_0-I^N_1&\overset{\xi_-=0}{\approx} 1 + \frac{U}{2},
\end{align}
\end{subequations}
and
\begin{subequations}
\begin{align}
I^N_0&\overset{\xi_-=2}{\approx}\dfrac{1}{\sqrt{2}}- \frac{3U}{4},
\\ 
I^N_1&\overset{\xi_-=2}{\approx}-\dfrac{1}{\sqrt{2}}-\frac{U}{2},
\\
I^N_0-I^N_1&\overset{\xi_-=2}{\approx}\sqrt{2}-\frac{U}{4},
\end{align}
\end{subequations}
respectively, which are in very good agreement with the values obtained numerically for
$U=5$ (see the bottom panels of Fig.~\ref{fig:IP_optgap_xim_w0}). Note
that, in the
latter case,
the EEXX optical gap equals
$\sqrt{2}-\frac{5}{4}\approx 0.16$ when $\xi_-=2$, which confirms the dramatic
underestimation of the optical gap that is observed in the bottom right panel of
Fig.~\ref{fig:IP_optgap_xim_w0} as $\xi_-$ increases.   





\newcommand{\Aa}[0]{Aa}

\end{document}